\pdfoutput=1

\documentclass[utf8]{frontiersSCNS} 

\usepackage{url,hyperref,lineno,microtype,float}
\hypersetup{colorlinks=true, urlcolor=blue}
\usepackage[onehalfspacing]{setspace}

\newcommand*{\defeq}{\stackrel{\text{def}}{=}}
\usepackage{multirow}
\usepackage{float}

\usepackage[bordercolor=white,backgroundcolor=gray!30,linecolor=black,colorinlistoftodos]{todonotes}

\DeclareFontFamily{OT1}{pzc}{}
\DeclareFontShape{OT1}{pzc}{m}{it}{<-> s * [1.200] pzcmi7t}{}
\DeclareMathAlphabet{\mathpzc}{OT1}{pzc}{m}{it}

\usepackage{subcaption}
\usepackage[labelformat=parens,labelsep=quad,skip=3pt]{caption}
\usepackage{graphicx}

\def\keyFont{\fontsize{8}{11}\helveticabold }
\def\firstAuthorLast{Côté-Allard, U., Campbell, E., {et~al.}} 
\def\Authors{Ulysse Côté-Allard\,$^{\dagger, 1, *}$, Evan Campbell\,$^{\dagger, 2, *}$, Angkoon Phinyomark\,$^{2}$,\\ François Laviolette\,$^{3}$, Benoit Gosselin\,$^{\ddagger,1}$ and Erik Scheme\,$^{\ddagger,2}$}
\def\Address{$\dagger$ These authors share first authorship \\ 
$\ddagger$ These authors share senior authorship\\
$^{1}$Department of Computer and Electrical Engineering, Université Laval, Quebec, QC, Canada \\
$^{2}$Department of Electrical and Computer Engineering, Institute of Biomedical Engineering, University of New Brunswick, Fredericton , NB, Canada\\
$^{3}$Department of Computer Science and Software Engineering, Université Laval, Quebec, QC, Canada}
\def\corrAuthor{Ulysse Côté-Allard}
\def\corrEmail{ulysse.cote-allard.1@ulaval.ca}

\begin{document}
\onecolumn
\firstpage{1}

\title[Handcrafted vs Deep Learning Features]{Interpreting Deep Learning Features for Myoelectric Control: A Comparison with Handcrafted Features}

\author[\firstAuthorLast ]{\Authors} 
\address{} 
\correspondance{} 

\extraAuth{Evan Campbell \\ evan.campbell1@unb.ca}
\maketitle
\begin{abstract}
Existing research on myoelectric control systems primarily focuses on extracting discriminative characteristics of the electromyographic (EMG) signal by designing handcrafted features. Recently, however, deep learning techniques have been applied to the challenging task of EMG-based gesture recognition. The adoption of these techniques slowly shifts the focus from feature engineering to feature learning. Nevertheless, the black-box nature of deep learning makes it hard to understand the type of information learned by the network and how it relates to handcrafted features. Additionally, due to the high variability in EMG recordings between participants, deep features tend to generalize poorly across subjects using standard training methods. Consequently, this work introduces a new multi-domain learning algorithm, named ADANN (Adaptive Domain Adversarial Neural Network), which significantly enhances ($p=0.00004$) inter-subject classification accuracy by an average of 19.40\% compared to standard training.

Using ADANN-generated features, this work provides the first topological data analysis of EMG-based gesture recognition for the characterisation of the information encoded within a deep network, using handcrafted features as landmarks. This analysis reveals that handcrafted features and the learned features (in the earlier layers) both try to discriminate between all gestures, but do not encode the same information to do so.
In the later layers, the learned features are inclined to instead adopt a one-versus-all strategy for a given class. Furthermore, by using convolutional network visualization techniques, it is revealed that learned features actually tend to ignore the most activated channel during contraction, which is in stark contrast with the prevalence of handcrafted features designed to capture amplitude information. Overall, this work paves the way for hybrid feature sets by providing a clear guideline of complementary information encoded within learned and handcrafted features.

\tiny
 \keyFont{ \section{Keywords:} EMG, Deep Learning, MAPPER, feature extraction, gesture recognition, CNN, ConvNet, t-SNE, Grad-CAM} 
\end{abstract}

\section{Introduction}
\label{introduction}

\quad Surface Electromyography (sEMG) is a technique employed in a vast array of applications from assistive technologies~\citep{prosthetics_EMG, wheelchairs_EMG} to bio-mechanical analysis~\citep{emg_deadlift}, and more generally as a way to interface with computers and robots~\citep{emg_game, emg_dance}. 
Traditionally, the sEMG-based gesture recognition literature primarily focuses on feature engineering as a way to increase the information density of the signal to improve gesture discrimination~\citep{emg_survey, state_of_the_art_review_emg_Erik,  features_selection_emg}. 
In the last few years, however, researchers have started to leverage deep learning~\citep{convNet_emg_1, robotic_arm_emg_control, deep_emg_review}, shifting the paradigm from feature engineering to feature learning.

Deep learning is a multi-level representation learning method (i.e. methods that learn an embedding from an input to facilitate detection or classification), where each level generates a higher, more abstract representation of the input~\citep{deep_learning_presentation}. 
Conventionally, the output layer (i.e., classifier or regressor) only has direct access to the output of the highest representation level~\citep{deep_learning_presentation, convNet_overview_history_from_alexnet}.
In contrast, several works have also fed the intermediary layers' output directly to the network's head~\citep{Pedestrian_convNet_using_intermediary_feature, all_intermediary_layers_to_last_layer, dagNet_NLP}. 
Arguably, the most successful approach using this design philosophy is DenseNet~\citep{denseNet}, a type of convolutional network (ConvNet) where each layer receives the feature maps of all preceding layers as input. 
Features learned by ConvNets were also extracted to be employed in conjunction with (or replace) handcrafted features when training conventional machine learning algorithms (e.g., support vector machine, linear discriminant analysis, decision tree)~\citep{sentiment_analysis_extract_deep_features, handCrafted_vs_nonHandcrafted_vision, fusing_latent_ConvNet_features_to_use_on_conventional_machine_learning_algorithm, sEMG_learned_feature_extraction}. 
Within the context of sEMG-based gesture recognition, deep learning was shown to be competitive with the current state of the art~\citep{transferLearning_journal_ulysse} and when combined with handcrafted features, to outperform it~\citep{sEMG_learned_feature_extraction}. 
This last result seems to indicate that, for sEMG signals, deep-learned features provide useful information that may be complementary to those that have been engineered throughout the years. However, the black box nature of these deep networks means that understanding what type of information is encapsulated throughout the network, and how to leverage this information, is challenging. 

The main contribution of this work is, therefore, to provide the first extensive analysis of the relationship between handcrafted and learned features within the context of sEMG-based gesture recognition. Understanding the feature space learned by the network could shed new insights on the type of information contained in sEMG signals. In turn, this improved understanding will allow the creation of better handcrafted features and facilitate the creation of new hybrid feature sets using this feature learning paradigm.

An important challenge arises when working with biosignals, as extensive variability exists between subjects~\citep{emg_running_inter_subject_variation, ecg_inter_subject_variation, eeg_inter_subject_variation, cross_subject_classification_normal_method, emg_normalzation_variation_between_subjects}. Especially within the context of sEMG-based gesture recognition~\citep{cross_subject_classification_normal_method, emg_normalzation_variation_between_subjects}. Consequently, features learned using traditional deep learning training methods can be highly participant-specific, which would hinder the goal of this work of learning a general feature representation of sEMG signals. By defining each participant as a different \textit{domain}, however, this issue can be framed as a Multi-Domain Learning problem (MDL)~\citep{multi_domain_learning_definition}, with the added restriction that the network's weights should be participant-agnostic. Multiple popular and effective MDL algorithms have been proposed over the years~\citep{multi_domain_learning, MDL_multiple_head, pre_train_MDL}. For example, \cite{MDL_multiple_head} proposed to use a shared network across multiples domains with one predictive head per domain. In~\cite{multi_domain_learning_definition}, a single head was shared across two parallel networks with one of them receiving the example's representation as input, while the other receives a vector representation of the associated domain of the example. These algorithms however are ill-suited for this work's context as they: do not explicitly impose domain-agnostic weight learning~\citep{multi_domain_learning_definition}, can scale poorly with the number of domains (i.e. participants)~\citep{MDL_multiple_head}, or are restricted to encode a single domain within their learned features (and use adaptor blocks to bridge the gap between domains)~\citep{pre_train_MDL}. Unsupervised domain-adversarial training algorithms~\citep{original_DANN, DANN, unshared_weights_domain_adversarial, dirt_t} predict an unlabeled dataset by learning a representation on a labeled dataset that makes it hard to distinguish between examples from either distribution. However, these algorithms are often not designed to learn a unique representation across more than two domains simultaneously~\citep{original_DANN, DANN, unshared_weights_domain_adversarial, dirt_t}, can be destructive to the source domain representation (through iterative process)~\citep{dirt_t}, and by nature of the problem they are trying to solve, do not leverage the labels of the target domains. As such, this work presents a new multi-domain adversarial training algorithm, named ADANN (Adaptive Domain Adversarial Neural Network). ADANN trains a network across multiple domains simultaneously while explicitly penalizing any domain-variant representations to study learned features that generalize well across participants. 

In this work, the sEMG information encapsulated within the general deep learning features learned by ADANN, is characterized using handcrafted features as landmarks in a topological network. This network is generated via the Mapper algorithm~\citep{mapper_original_article}, with \textit{t}-Stochastic Neighbor Embedding (t-SNE)~\citep{tSNE}, a non-linear dimensionality reduction visualization method, as the filter function.
Mapper is a Topological Data Analysis (TDA) tool that excels at determining the shape of high dimensional data, by providing a faithful representation of it through a topological network. 
This TDA tool has been applied as a solution to numerous challenging applications across a wide array of domains; for example, uncovering the dynamic organization of brain activity during various tasks~\citep{mapper_brain_organization} or identifying a subgroup of breast cancer with 100\% survival rate and no metastasis~\citep{mapper_breast_cancer}. 
Mapper has also been applied to determine relationships between feature space for physiological signal pain recognition~\citep{evan_pain}, and EMG-based gesture recognition~\citep{toon_navigatingfeatures}. However, to the best of the authors' knowledge, the use of TDA to interpret information harnessed within deep-learned features using handcrafted features as landmarks has yet to be explored. 

In this paper, convNet visualization techniques are also leveraged as a way to highlight how the network makes class-discriminant decisions. 
Several works~\citep{saliancy_maps, saliency_map_example_paper, deconvolution, guided_backpropagation} have proposed to visualize network's predictions by emphasizing which input-pixels have the most impact on the network's output, consequently, fostering a better understanding of what the network has learned. 
For example, \cite{saliancy_maps} used partial derivatives to compute pixel-relevance for the network output. Another example is Guided Backpropagation~\citep{guided_backpropagation}, which modifies the computation of the gradient to only include paths within the network that positively contribute to the prediction of a given class. When compared with saliency maps~\citep{saliancy_maps}, Guided Backpropagation results in qualitative visualization improvements~\citep{guided_grad_cam}. 
While these methods produce resolutions at a pixel level, the images produced with respect to different classes are nearly identical~\citep{guided_grad_cam}. 
Other types of algorithms provide highly class-discriminative visualizations, but at a lower resolution~\citep{cam, grad_cam} and sometimes require a specific ConvNet architecture~\citep{cam} to use. Within this work, Guided Gradient-weighted Class Activation Mapping (Guided Grad-CAM)~\citep{guided_grad_cam} is employed as it provides pixel-wise input resolution while being class-discriminative. Another advantage of this technique is that it can be implemented on any ConvNet-based architecture without requiring re-training. To the best of the authors' knowledge, this is the first time that deep learning visualization techniques are applied to EMG signals.

\section{Material and Methods}
\label{material_and_methods}

\begin{figure}[h!]
\begin{center}
\includegraphics[width=\textwidth]{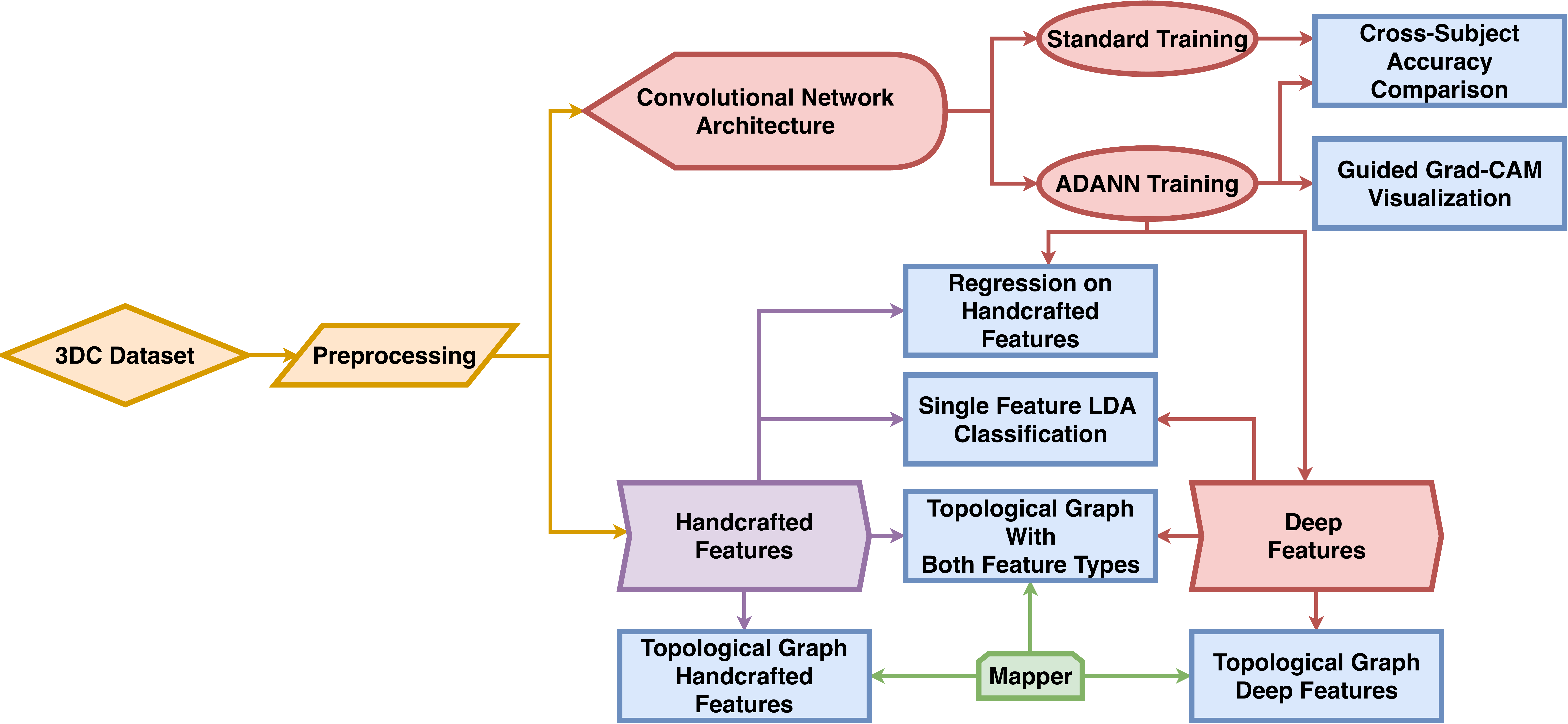}
\end{center}
\caption{Diagram of the workflow of this work. The \textit{3DC Dataset} is first preprocessed before being used to train the network using standard training and the proposed ADANN training procedure. The handcrafted features are directly calculated from the preprocessed dataset, while the deep features are extracted from the ConvNet trained with ADANN. In the diagram, the blue rectangles represent experiments and the arrows show which methods/algorithms are required to perform them.} \label{fig:flowchart_material_and_method}
\end{figure}

\quad A flowchart of the material, methods and experiment is shown in Figure~\ref{fig:flowchart_material_and_method}. This section is divided as follows: first, a description of the dataset and preprocessing used in this work is given in Section~\ref{dataset_section}. Then, the handcrafted features are presented in Section~\ref{hand_craft_features_section}. The ConvNet architecture and the new multi-domain adversarial training algorithm (ADANN) are presented in Section~\ref{ConvArchitecture} and \ref{multi_domain_adversarial_training_section}, respectively. A brief overview of Guided Grad-CAM is given in Section~\ref{guided_gradcam_explanation}, while Section~\ref{lda_training_explanation} and~\ref{RegressionModelSection} present single feature classification and handcrafted feature regression, respectively. Finally, the Mapper algorithm is detailed in Section~\ref{MapperSection}.

\subsection{EMG Data}
\label{dataset_section}

\begin{figure}[h!]
\begin{center}
\includegraphics[width=\textwidth]{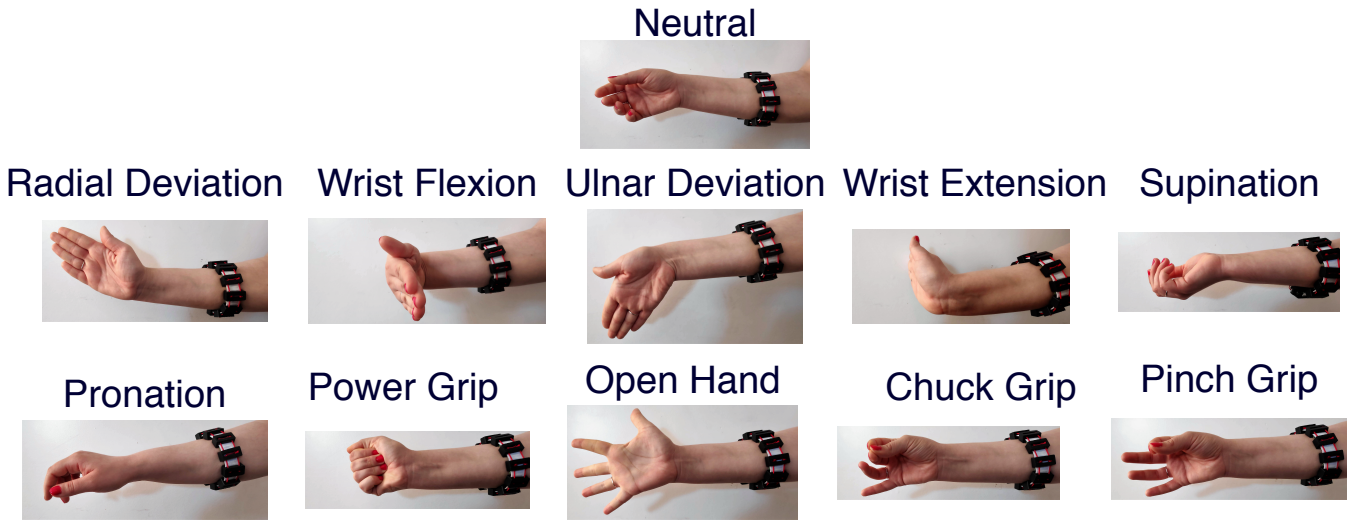}
\end{center}
\caption{The eleven hand/wrist gestures recorded in the \textit{3DC Dataset} (image re-used from~\citep{3DC_armband_and_dataset})} 
\label{fig:1_gestures_dataset}
\end{figure}

The dataset employed in this work is the \emph{3DC Dataset}~\citep{3DC_armband_and_dataset}, featuring 22 able-bodied participants performing ten hand/wrist gestures + neutral (see Figure~\ref{fig:1_gestures_dataset} for the list of gestures). 
This dataset was recorded with the 3DC Armband; a wireless, 10-channel, dry-electrode, 3D printed sEMG armband. The device samples data at 1000~Hz per channel, allowing the feature extraction to take advantage of the full spectra of sEMG signals~\citep{emg_200_vs_1000Hz}. Informed consent was obtained from all participants, as approved by Laval University's Research Ethics Committee \citep{3DC_armband_and_dataset}.

The dataset was built as follows: Each participant was asked to perform and hold each gesture for a period of five seconds starting from the neutral position to produce a \textit{cycle}. Three more cycles were recorded to serve as the \textit{training dataset}. After a five minute break, four new cycles were recorded to serve as the \textit{test dataset}. Note that the validation set and hyperparameter selection are made from the training dataset. 


As this work aims to understand the type of features learned by deep network in the context of myoelectric control systems, a critical factor to consider is the input latency. 
\cite{optimal_latency_real_time_EMG} showed that the optimal guidance latency was between 150 and 250~ms. As such, the data from each participant was segmented into 151~ms frames with an overlap of 100~ms. 
The raw data was then band-pass filtered between 20-495~Hz using a 4th-order Butterworth filter.

\subsection{Handcrafted Features}
\label{hand_craft_features_section}
Handcrafted features are characteristics extracted from windows of the EMG signal using established mathematical equations. The purpose of these feature extraction methods is to enhance the information density of the signal so as to improve discrimination between motion classes~\citep{emg_survey, features_selection_emg}.
Across the myoelectric control literature, hundreds of handcrafted feature extraction methods have been presented~\citep{emg_survey, features_selection_emg, 50_features_comparison_selection_emg}. As such, implementing the exhaustive set of features that has been proposed is impractical. Instead, within this study a comprehensive subset of 79 of the most commonly used features is employed. 
With a comprehensive set of features, past literature has identified five functional groups that summarize all sources of information current handcrafted feature extraction techniques describe: signal amplitude and power (SAP), nonlinear complexity (NLC), frequency information (FI), time-series modeling (TSM), and unique (UNI) \citep{toon_navigatingfeatures, evan_ner2019}. 
The SAP functional group includes time-domain energy or power features (e.g. Root Mean Squared, Mean Absolute Value).
The FI functional group generally refers to features extracted from the frequency domain, or features that describe spectral properties (e.g. Mean Frequency, Zero Crossings). 
The NLC functional group corresponds to features that describe entropy or similarity based information (e.g. Sample Entropy, Maximum Fractal Length). 
The TSM functional group represents features that attempt to reconstruct the data provided through stochastic or other algorithmic models (e.g. Autoregressive Coefficients, Cepstral Coefficients). 
Finally, the UNI functional group represents features that capture various other modalities of information, such as measures of signal quality or a combination of other functional groups (e.g. Signal to Motion Artefact Ratio, Time Domain Power Spectral Descriptors).

\begin{table}[!htb]
\centering
\resizebox{\textwidth}{250pt}{%
\begin{tabular}{llll}
\textbf{Ref} & \textbf{Feature Extraction Method}  & \textbf{Name} & \textbf{Group}\\ \hline%
\citep{features_selection_emg}    & Amplitude of the First Burst                  & AFB     &   SAP \\
\citep{damv}    & Difference Absolute Mean Value                & DAMV    &   SAP \\
\citep{damv}    & Difference Absolute Standard Deviation Value  & DASDV   &   SAP \\
\citep{ar}    & Difference Log Detector                       & DLD     &   SAP \\
\citep{features_selection_emg}    & Difference Temporal Moment                    & DTM     &   SAP \\
\citep{ar}    & Difference Variance Value                     & DVARV   &   SAP \\
\citep{ar}    & Difference v-Order                            & DV      &   SAP \\
\citep{ar3}    & Integral of Electromyogram                    & IEMG    &   SAP \\ 
\citep{ar}    & Log Detector                                  & LD      &   SAP \\ 
\citep{m2}    & Second-Order Moment                           & M2      &   SAP \\
\citep{fr3}    & Modified Mean Absolute Value 1                & MMAV1   &   SAP \\ 
\citep{fr3}    & Modified Mean Absolute Value 2                & MMAV2   &   SAP \\ 
\citep{mav}    & Mean Absolute Value                           & MAV     &   SAP \\ 
\citep{features_selection_emg}    & Maximum                                       & MAX     &   SAP \\
\citep{mhw}    & Multiple Hamming Windows                      & MHW     &   SAP \\
\citep{mhw}    & Mean Power                                    & MNP     &   SAP \\
\citep{mhw}    & Multiple Trapezoidal Windows                  & MTW     &   SAP \\
\citep{mav}    & Root Mean Squared                             & RMS     &   SAP \\
\citep{mhw}    & Spectral Moment                               & SM      &   SAP \\
\citep{mhw}    & Sum of Squared Integral                       & SSI     &   SAP \\
\citep{features_selection_emg}    & Temporal Moment                               & TM      &   SAP \\
\citep{mhw}    & Total Power                                   & TTP     &   SAP \\
\citep{ar}    & Variance                                      & VAR     &   SAP \\
\citep{ar}    & v-Order                                       & V       &   SAP \\
\citep{features_selection_emg}    & Waveform Length                               & WL      &   SAP \\ \hline

\citep{fr2,fr3}    & Frequency Ratio                               & FR      &   FI \\
\citep{mnf,mnf2}    & Median Frequency                              & MDF     &   FI \\
\citep{mnf,mnf2}    & Mean Frequency                                & MNF     &   FI \\
\citep{features_selection_emg}    & Slope Sign Change                             & SSC     &   FI \\
\citep{ar}    & Zero Crossings                                & ZC      &   FI \\ \hline

\citep{apen}    & Sample Entropy                                & SAMPEN  &   NLC \\
\citep{apen}    & Approximate Entropy                           & APEN    &   NLC \\
\citep{ar}    & Willison's Amplitude                          & WAMP    &   NLC \\
\citep{bc}    & Box-Counting Fractal Dimension                & BC      &   NLC \\
\citep{katz2}    & Katz Fractal Dimension                        & KATZ    &   NLC \\
\citep{hg2}    & Maximum Fractal Length                        & MFL     &   NLC \\
\hline

\citep{ar3}    & Autoregressive Coefficients                   & AR      &   TSM \\
\citep{ar3}    & Cepstral Coefficients                         & CC      &   TSM \\
\citep{ar3}    & Difference Autoregressive Coefficient         & DAR     &   TSM \\
\citep{ar3}    & Difference Cepstral Coeffients                & DCC     &   TSM \\
\citep{dfa,dfa2}    & Detrend Fluctuation Analysis                  & DFA     &   TSM \\
\citep{psr}    & Power Spectrum Ratio                          & PSR     &   TSM \\
\citep{dpr,dpr2}    & Signal to Noise Ratio                         & SNR     &   TSM \\ \hline

\citep{ce,ce2}    & Critical Exponent                             & CE      &   UNI \\
\citep{dpr,dpr2}    & Maximum to Minimum Drop in Power Density Ratio& DPR     &   UNI \\
\citep{features_selection_emg}    & Histogram                                     & HIST    &   UNI \\
\citep{kurt,kurt2}    & Kurtosis                                      & KURT    &   UNI \\
\citep{features_selection_emg}    & Mean Absolute Value Slope                     & MAVS    &   UNI \\
\citep{dpr,dpr2}    & Power Spectrum Deformation                    & OHM     &   UNI \\
\citep{apen}    & Peak Frequency                                & PKF     &   UNI \\
\citep{psdfd}    & Power Spectrum Density Fractal Dimension      & PSDFD   &   UNI \\
\citep{kurt,kurt2}    & Skewness                                      & SKEW    &   UNI \\
\citep{dpr,dpr2}    & Signal to Motion Artefact Ratio               & SMR     &   UNI \\
\citep{m2}    & Time Domain Power Spectral Descriptors        & TSPSD   &   UNI \\
\citep{features_selection_emg}    & Variance of Central Frequency                 & VCF     &   UNI \\
\citep{apen}    & Variance Fractal Dimension                    & VFD     &   UNI \\ \hline

\end{tabular}}
\caption{Handcrafted features extracted for topological landmarks sorted by functional group. }\label{table:EMGFeatures}
\end{table}

Table~\ref{table:EMGFeatures} presents the 56 handcrafted feature methods considered in this work. Note that some methods produce multiple features (e.g. Cepstral Coefficients, Histogram), resulting in a total of 79 features.
The SAP, FI, NLC, TSM, and UNI feature groups are represented here by 25, 5, 6, 7, and 13 feature extraction methods respectively. 
In the TDA of the deep learned features (see Section~\ref{MapperSection}), these handcrafted features serve as landmarks for well-understood properties of the EMG signal. 
In the regression model analysis (see Section~\ref{RegressionModelSection}), the flow of information through the ConvNet is visualized by employing the handcrafted features methods as the target of the network.


\vspace*{1cm}

\subsection{Convolutional Network}

The following subsections present the deep learning architecture, training methods and visualization techniques employed in this paper. 
The PyTorch ~\citep{pytorch} implementation employed in this work is \href{https://github.com/UlysseCoteAllard/sEMG\_handCraftedVsLearnedFeatures}{available here}.

\subsubsection{Architecture}
\label{ConvArchitecture}

Recent works on sEMG-based gesture recognition using deep learning have shown that ConvNets trained with the raw sEMG signal as input were able to achieve similar classification accuracy to the current state of the art~\citep{raw_emg_as_input_convNet_very_good, transferLearning_journal_ulysse}. 
Consequently, and to reduce bias, the preprocessed raw data (see Section~\ref{dataset_section}) is passed directly as an image of shape \textbf{10 $\times$ 151} (Channel $\times$ Sample) to the ConvNet. 

\begin{figure}[h!]
\begin{center}
\includegraphics[width=\textwidth]{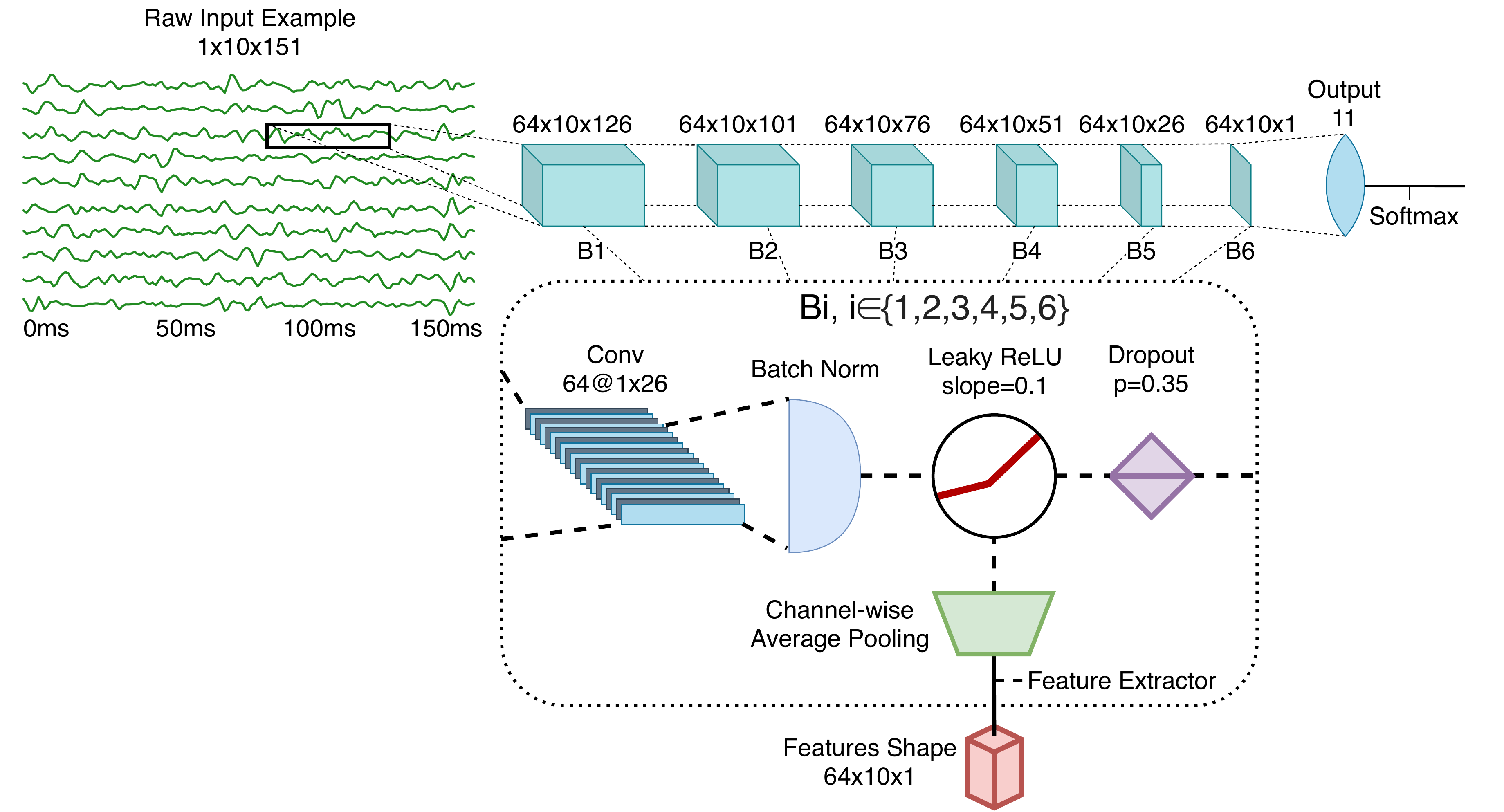}
\end{center}
\caption{The ConvNet's architecture, employing 543,629 learnable parameters. In this figure, \textit{Bi} refers to the ith feature extraction block (i$\in$\{1,2,3,4,5,6\}). Conv refers to Convolutional layer. As shown, the feature extraction is performed after the non-linearity (leaky ReLU).} \label{fig:2_ConvNet_architecture}
\end{figure}

The ConvNet's architecture, which is depicted in Figure~\ref{fig:2_ConvNet_architecture}, contains six \textit{blocks} followed 
by a fully connected layer for gesture-classification. The network's topology was selected to obtain a deep network with a limited number of learnable parameters (to avoid overfitting) with simple layer connections to enable an easier, and thus more thorough analysis. All architecture choices and hyperparameter selection were performed using the training set of the \textit{3DC Dataset} or inspired by previous works (\citep{transferLearning_journal_ulysse} and~\citep{3DC_armband_and_dataset}).
Each block encapsulates a convolutional layer~\citep{deep_learning_presentation}, followed by batch normalization (BN)~\citep{batch_normalization}, leaky ReLU (slope=0.1)~\citep{leaky_relu} and dropout~\citep{dropout} (with a drop rate set at 0.35 following~\cite{transferLearning_journal_ulysse}).  
The number of blocks within the network was selected to obtain a sufficiently deep network to study how the type of learned features evolve with respect to their layer. The depth of the network was limited by the number of examples available for training and more complex layer connections (e.g. residual network~\citep{resnet}, dense network~\citep{denseNet}) were avoided to not ambiguate the analysis performed in this work. The number of feature maps (64) was kept uniform for each layer, allowing for easier comparisons of learned features across the convolutional layers. 
The filter size was \textbf{1 $\times$ 26} so that, similarly to the handcrafted features, the learned features are channel independent. Due to the selected filter size, the dimensions of feature maps at the final layer is 10 $\times$ 1.

Adam~\citep{adam} was employed to optimize the ConvNet with an initial learning rate of 0.0404709 and batch size of 512 (as used in~\citep{3DC_armband_and_dataset}). 
The training dataset was divided into training and validation sets using the first three cycles and last cycle, respectively. 
Employing this validation set, learning rate annealing was applied with a factor of five and a patience of fifteen with early stopping applied when two consecutive annealings occurred without achieving a better validation loss.

For the purpose of the TDA, features maps were extracted after the non-linearity using per feature-map channel-wise average pooling. That is, the number of feature maps remained the same, but the feature map's value per channel was averaged to a single scalar (as is common with handcrafted features).

\subsubsection{Multi-Domain Adversarial Training}
\label{multi_domain_adversarial_training_section}

To better understand what type of features are commonly learned at each layer of the network, it is desirable that the model generalizes well across participants. This feature generality principle also motivates the design of the handcrafted features (presented in Section~\ref{hand_craft_features_section}), as it would be impractical to create new features for each new participant. Learning a general feature representation across participants, however, cannot be achieved by simply aggregating the training data of all participants and then training a classifier normally. As, even when precisely controlling for electrode placement, cross-subject accuracy using standard learning methods is poor~(\cite{cross_subject_classification_normal_method}). This problem is compounded by the fact that important differences exist between subjects of the \textit{3DC Dataset} (i.e. position and rotation of the armband placed on the left or right arm).

Learning a participant-agnostic representation can be framed as a multi-domain learning problem~\citep{multi_domain_learning}. In the context of sEMG-based gesture recognition, AdaBN, a domain adaptation algorithm presented in~\citep{adabatch}, was successfully employed as a way to learn a general representation across participants in~\citep{conference_transfer_learning_algorithm, transferLearning_journal_ulysse}. The hypothesis of AdaBN is that label-related information (i.e. hand gestures) will be contained within the network's weights, while the domain-related information (i.e. participants) are stored in their BN statistics. Training is thus performed by sharing the weights of the network across the subjects dataset while tracking the BN statistics independently for each participant. 

\begin{figure}[h!]
\begin{center}
\includegraphics[width=\textwidth]{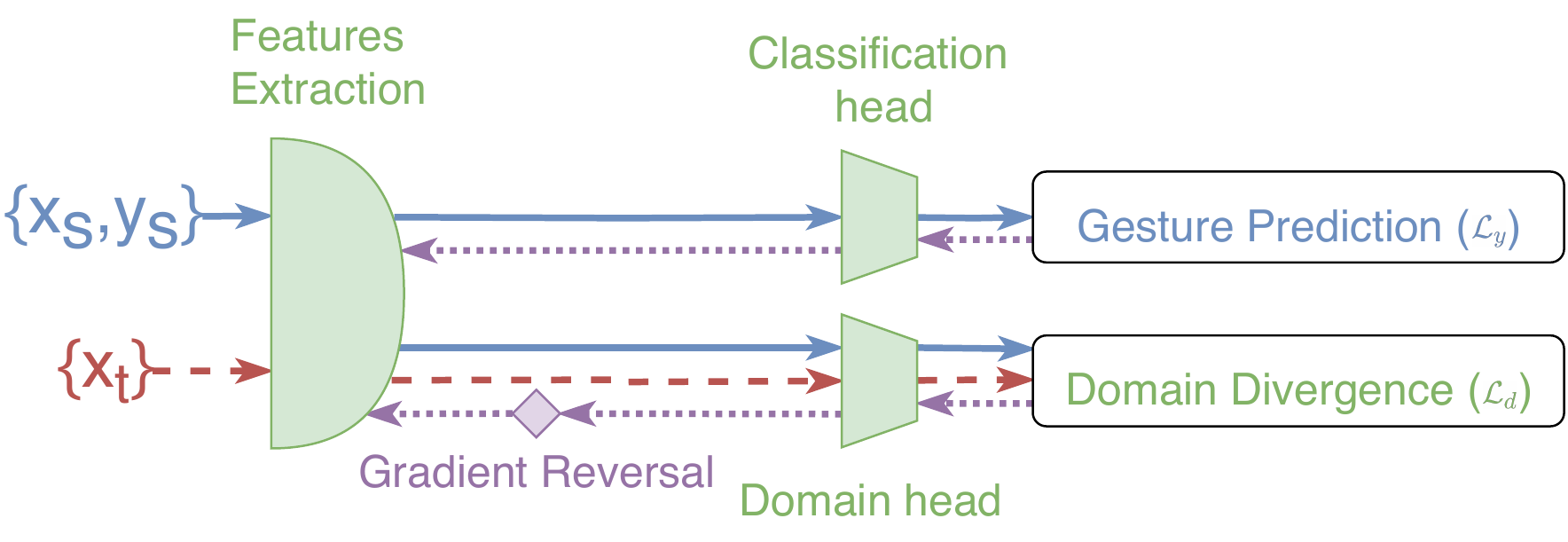}
\end{center}
\caption{Overview of the training steps of ADANN (identical to DANN) for one labeled batch from the source (\{$x_s, y_s$\}, blue lines) and one unlabeled batch from the target (\{$x_t$\}, red dashed lines). The purple dotted lines correspond to the backpropagated gradient. The gradient reversal operation is represented by the purple diamond.}
\label{fig:ADANN_algorithm}
\end{figure}

To inhibit the shared network's weights from learning subject-specific representation, Domain-Adversarial Neural Networks (DANN) training~\citep{DANN} is employed. DANN is designed to learn domain-invariant features across two domains from the point of view of the desired task. The approach used by DANN to achieve this objective consists of adding a second head (referred to as the \textit{domain classification head}) to the network presented in Section~\ref{ConvArchitecture}, which receives the output of block B6. The goal of this second head is to learn to discriminate between the domains. However, during backpropagation, the gradient computed from the domain loss is multiplied by a negative constant (set to -1 in this work) as it exits the domain classification head. This gradient reversal explicitly forces the feature distributions over the domains to be similar. Note that the backpropagation algorithm proceeds normally for the first head (gesture classification head). The loss function used for both heads is the cross-entropy loss. The two losses are combined as follows: $\mathcal{L}_y + \lambda\mathcal{L}_d$, where $\mathcal{L}_y$ and $\mathcal{L}_d$ are the prediction and domain loss, respectively (see Figure~\ref{fig:ADANN_algorithm}), while $\lambda$ is a scalar that weights the domain loss (set to 0.1 in this work).

Using this approach, each participant of the 3DC Dataset represents a different domain (n=22). A direct application of DANN would thus initialize the domain classification head with 22 output neurons. This, however, could create a pitfall where the network is able to differentiate between the domains perfectly while simply predict one of the 21 other domains to maximize $\mathcal{L}_d$. Instead, the domain classification head is initialized with only two output neurons. At each epoch, a batch is created that contains examples from a single participant (this batch is referred to as the \textit{source batch}, and is assigned the domain label 0). A second batch, referred to as the \textit{target batch}, is also created that contains examples from one of the other participants selected at random, and is assigned the domain label 1. As every participants data is used as the source batch at each epoch, this ensures that the network is forced to learn a domain-independent feature representation. ADANN's goal is thus to force the network to be unable to accurately associate a participant with their examples while achieving a highly discriminative gesture representation across all participants. During training, the BN statistics are tracked individually for each subject. Therefore, when learning from a source or target batch, the network uses the BN statistics associated with the corresponding participant. Note that, by construction, the participant associated with the source is necessarily different from the participant associated with the target. Consequently, the network is fed the source and target batch consecutively (i.e. not both batch simultaneously). Also note that the BN statistics are updated only in association with the source batch to ensure equal training updates across all participants. For a given iteration, once the source and target batch are constructed, the training step proceeds as described for DANN (see Figure~\ref{fig:ADANN_algorithm}).

To assess the performance of the proposed MDL algorithm, two identical ConvNet (as described in Section~\ref{ConvArchitecture}) were created. One of the ConvNets was trained with ADANN, whereas the other used a standard training loop (i.e., aggregating the data from all participants), with both using the same hyperparameters. The networks trained with both methods were then tested on the test dataset with no participant-specific fine-tuning.

\vspace*{1cm}

\subsubsection{Learning Visualization}
\label{guided_gradcam_explanation}

One of the main problems associated with deep learning is interpretability of how and why a model makes a prediction given a particular input. A first step in understanding a network prediction is through the visualization of the learned weights, feature maps and gradients resulting from a particular input. Consequently, several sophisticated visualisation techniques have been developed, which are aimed at facilitating a better comprehension of the hierarchical learning that takes place within a network~\citep{saliancy_maps, guided_backpropagation, cam}. One popular such technique is Guided Grad-CAM, which combines high resolution pixel-space gradient visualization and class-discriminative visualization~\citep{guided_grad_cam}. Guided Grad-CAM is thus employed to visualize how the ConvNet trained with ADANN makes its decisions, both on real examples from the \textit{3DC Dataset} and on an artificially generated signals. 

Given an image that was used to compute a forward pass in the network and a label $y$, the output of Guided Grad-CAM is calculated from four distinct steps (note that steps two and three are computed independently from each other using the output of step one):

\begin{enumerate}
\item Set all the gradients of the output neurons to zero, except for the gradient of the neuron associated with the label $y$ ( which is set to one) and name the gradient of the neuron of interest $y^g$.
\item  Set all negative activations to zero. Then, perform backpropagation, but before propagating the gradient at each step, set all the negative gradients to zero again. Save the final gradients corresponding to the input image. 
This step corresponds to computing the guided backpropagation~\citep{guided_backpropagation}.
\item Let $F_{j,i}$ be the activation of the $i$th feature map of the $j$th layer with feature maps of the network. Select a layer $F_j$ of interest (in this work $F_j$ correspond to the rectified convolutional layer of B6). Backpropagate the signal from the output layer to $F_{j, i}$ (i.e. $\frac{\partial y^g}{\partial F_{j,i}}$). Then for each $i$ compute the global average pooling of $\frac{\partial y^g}{\partial F_{j,i}}$ and name it $w_{j,i}$. Finally, compute:
$ReLU\left(\sum_i w_{j,i} F_{j,i}\right)$

This third step corresponds to computing the Gradient-weighted Class Activation Mapping (Grad-CAM)~\citep{grad_cam}.
\item Finally, fuse the output of the two previous steps using point-wise multiplication to obtain the output of Guided Grad-CAM~\citep{guided_grad_cam}.
\end{enumerate}

\vspace*{1cm}

\subsubsection{Learned feature classification}
\label{lda_training_explanation}
Similarly to~\cite{sEMG_learned_feature_extraction}, the learned features were extracted to train a Linear Discriminant Analysis (LDA) classifier to show the discriminative ability of the learned features. LDA was selected as it was shown to provide robust classification within the context of sEMG-based gesture recognition~\citep{Campbell2019SIP}, does not require hyperparameter tuning, and creates linear boundaries within the input feature space.
LDA was trained in a cross-subject framework on the training dataset and tested on the test dataset. For comparison purposes, LDA was also trained on the handcrafted features described in Section~\ref{hand_craft_features_section}. Note that the implementation was from scikit-learn~\citep{scikit_learn}. 

\vspace*{1cm}

\subsubsection{Regression Model}
\label{RegressionModelSection}
One method of highlighting the information content encoded throughout a network is to see how well known handcrafted features can be predicted from the network's feature maps at different stages. This can be achieved using an added output neuron (\textit{regression head}) at the feature extraction stage (i.e. after the non-linearity, but before the average pooling (before the green trapezoid of Figure~\ref{fig:2_ConvNet_architecture})) of each block.  The goal of this output is to map from the learned features to the handcrafted features of interest. 
As all the features considered in Section~\ref{hand_craft_features_section} are calculated channel-wise, only the information from the first sEMG channel (arbitrarily selected) of the feature maps will be fed to the regression head.

The training procedure to implement this is as follows: first, pre-train the network using ADANN (presented in Section~\ref{multi_domain_adversarial_training_section}). Second, freeze all the weights of the network, except for the weights associated with the regression head of the block of interest. The Mean Square Error (MSE) is then employed as the loss function with the target being the value of the handcrafted feature of interest from the first sEMG channel. Due to the stochastic nature of the algorithm, the training was performed 20 times for each participant and the results were given as the average MSE computed on the test dataset across of all participants. Note that the targets derived from multi-output feature extraction methods (e.g. Autoregressive Coefficients) corresponded to the first principal component returned by Principal Component Analysis (PCA) (where singular value decomposition was performed on the training and test set for the training and test phase, respectively).

\vspace*{1cm}

\subsection{Topological Data Analysis - Mapper}
\label{MapperSection}




Conventional TDA methods such as Isomap~\citep{isomap} produce a low dimensional embedding by retaining geodesic distances between neighboring points. However, they often have limited topological stability~\citep{kernel_isomap} and lack the ability to produce a simplicial complex (a ball-and-stick simplification of the shape of the dataset) with size smaller than the original dataset~\citep{singh_originalmapper}. The Mapper algorithm~\citep{singh_originalmapper} is a TDA method that creates interpretable simplifications of high-dimensional data sets that remain true to the shape of the data set.
Mapper can thus produce a stable representation of the topological shape of the dataset at a specified resolution, where the shape of the network has been simplified during a partial clustering stage.
Further, the shape of the dataset is defined such that it is coordinate, deformation, and compression invariant. Consequently, this TDA algorithm can be employed to better understand how handcrafted and deep-learned features relate to one-another. In this work, Mapper is employed on three scenarios; (A), (B) and (C). In scenario (A), the algorithm only uses the handcrafted features as a way to validate the hyperparameters selected by cross-referencing the results with previous EMG works using Mapper~\citep{toon_navigatingfeatures, evan_ner2019}. For scenario (B), only the learned features are used to determine if features within the same block extract similar or dissimilar sources of information (i.e. the degree at which the features within the same block are dispersed across the topological network). Finally, in scenario (C), Mapper is applied to the combination of learned and handcrafted features to better understand their relationship and to provide new avenues of research for sEMG-based gesture recognition. 

Sections \ref{mapper_algo} through \ref{mapper_math}, below, provide additional details about the approach, mathematical basis and implementation of Mapper in this work. Readers who are familiar with, or prefer to avoid these details, may jump directly to Section \ref{results}. 

\vspace*{0.5cm}

\subsubsection{Mapper Algorithm}
\label{mapper_algo}

The construction of the topological network created using the Mapper algorithm can be seen as a five stage pipeline:
\begin{enumerate}
    \item \textit{prepare}: organize the data set to produce a point cloud of features in high dimensional space.
    \item \textit{lens}: filter the high dimensional data into a lower dimensional representation using a lens.
    \item \textit{resolution}: divide the filtration into a set of regions.
    \item \textit{partial clustering}: for each region, cluster the contents in the original high dimensional space.
    \item \textit{combine}: combine the region isolated clusters into a single topological network using common points across regions~\citep{DyNeSR}.
\end{enumerate}

\subsubsection{Mathematical definition of Mapper}
\label{mapper_math}

A mathematical definition of the Mapper algorithm for feature extraction using a multi-channel recording device is as follows:

Let $\mathbf{x} \defeq (\vec{x}_1, ..., \vec{x}_C)$ be a series of samples for each $C$ channels, where $\vec{x}_c \in \mathbb{R}^S, \forall c\in\{1, ...,C\}$ and $S$ is the length of a consecutive series of data. Define $\mathpzc{X}\defeq \{\mathbf{x}_n\}_{n=1}^N$ a set of $N$ examples. Let also $\Phi \defeq \{\phi_m\}^M_{m=1}$ be a set of $M$ feature-generating functions of the form $\phi_m: \mathbb{R}^S\rightarrow\mathbb{R}$. Given $\mathbf{x}_{n, c}$ the $c$~th element of $\mathbf{x}_n\in\mathpzc{X}$, the resulting feature $f_{n,c}^m\in\mathbb{R}$ is obtained by applying $\phi_m$ such that $f_{n,c}^m\defeq\phi_m(\mathbf{x}_{n, c})$. Consequently, the vector $\vec{f}_m\in\mathbb{R}^{N\times C}$ is obtained such that $ \vec{f}_m \defeq (f_{1,1}^m, f_{1,2}^m, ..., f_{1,C}^m, f_{2,1}^m, f_{2,2}^m, ..., f_{2,C}^m ,...,f_{N, C}^m)$. 


The first step of the Mapper algorithm is to consider $\mathpzc{F}\defeq \{\vec{f}_m\}_{m=1}^M$, the transformed data points from $\mathpzc{X}$. Then define $\psi: \mathbb{R}^{N \times C}\rightarrow\mathbb{R}^{Z}$, with $0 < Z \ll N \times C$ and consider the set $\mathpzc{Z} \defeq \{\psi(\vec{f}) | \vec{f}  \in \mathpzc{F}\}$. This dimensionality reduction ($N\times C\rightarrow Z$) is employed to reduce the computational cost of the rest of the Mapper algorithm and can be considered as a hyperparameter of the Mapper algorithm. 


In the second step of the algorithm, define $\sigma: \mathbb{R}^{Z}\rightarrow\mathbb{R}^{W}$, with $0 < W \ll Z$ and consider the set $\mathpzc{W}\defeq\{\sigma(\vec{z}) | \vec{z} \in \mathpzc{Z}\}$.
In the literature~\citep{mapper_original_article}, the function $\sigma$ is called filter function and $\mathpzc{W}$ is the image or lens.

\begin{figure}[h!]
\begin{center}
\includegraphics[width=\textwidth]{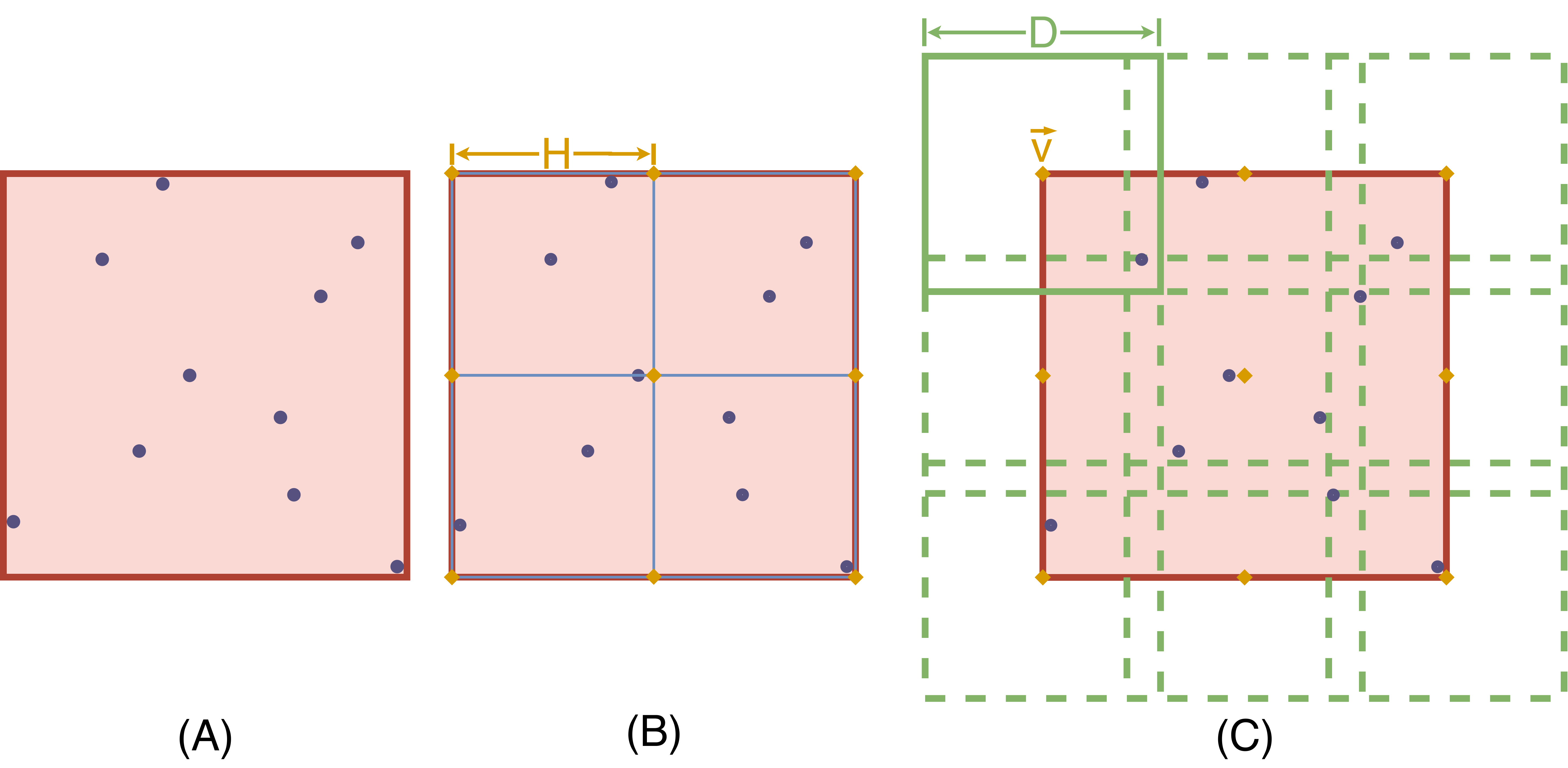}
\end{center}
\caption{An example of step 3 of the Mapper algorithm with $W=2$. The purple dots represent the elements of $\mathpzc{W}$. In (A), the red square corresponds to $\mathfrak{C}$. In (B), $\mathfrak{C}$ is subdivided using $k^2$ squares of length $H$ (with $k=2$ in this case). The orange diamonds, in both (B) and (C), represent the elements of $\mathpzc{V}$. Finally, the square $\mathpzc{c}_{\vec{v}}$ of length $D$ is shown on the upper left corner of (C), overlapping other squares centered on other elements of $\mathpzc{V}$ (dotted lines).}
\label{fig:mapper_bin_example}
\end{figure}

Third, let $\mathfrak{C}$ be the smallest hypercube of $\mathbb{R}^W$ which covers $\mathpzc{W}$ entirely. As $\mathpzc{X}$ is a finite set, each dimension of $\mathfrak{C}$ is a finite interval. Let $k\in\mathbb{N}^*$, be a hyperparameter that subdivides $\mathfrak{C}$ evenly into $k^W$ smaller hypercubes. Note that the side lengths of these smaller hypercubes are $H=\frac{1}{k}\times$ the length size of $\mathfrak{C}$. Denotes $\mathpzc{V}$ the set of all vertices of these smaller hypercubes. Next, fix $D>H$ as another hyperparameter. For each $\vec{v}\in\mathpzc{V}$, consider the hypercube $\mathpzc{c}_{\vec{v}}$ of length $D$ centered on $\vec{v}$. A visualization of step 3 is given in Figure~\ref{fig:mapper_bin_example}.  

Fourth, define $\mathpzc{Z}_{\vec{v}}\defeq \{\vec{z}\in\mathpzc{Z}| \sigma(\vec{z})\in\mathpzc{c}_{\vec{v}}\}$, the set of all elements of $\mathpzc{Z}$ that is projected in the hypercube $\mathpzc{c}_{\vec{v}}$. Let $\xi$ be a clustering algorithm and $\xi(\mathpzc{Z}_{\vec{v}})$ be the resulting set of clusters. Define $\mathpzc{B}$ as the set that consist of all so obtained clusters for all $\mathpzc{Z}_{\vec{v}}$.


Fifth, compute the topological graph $\mathpzc{G}$ using each element of $\mathpzc{B}$ as a vertex and create an edge between vertices $\mathpzc{G}_i$ and $\mathpzc{G}_j$ ($i,j\in\{1, ...,|\mathpzc{B}|\},i\neq j$) if
$\mathpzc{G}_i \cap \mathpzc{G}_j \neq \emptyset$.

\subsubsection{Mapper implementation within this work}
\label{mapper_imp}
In this work, as described in Section~\ref{dataset_section} the dataset was recorded using the 3DC Armband which offers 10 channel-recording ($C$=10) and an example is comprised of 151 data-points ($S$=151) for each channel. The number of considered features in scenarios (A), (B) and (C), are 79, 384, and 465, respectively. Note that multi-output feature extraction techniques (e.g. AR, HIST), consider each component of that vector as a separate feature. Each element of $\mathpzc{F}$ is obtained by computing the result of a feature from Section~\ref{hand_craft_features_section} (corresponding to $\phi_m()$ in the mathematical definition given previously) over each channel of each example of the \textit{Training Dataset}. The dataset undergoes the first dimensionality reduction ($\Psi()$) using PCA~\citep{PCA}, where the number of principal components used corresponds to 99\% of the total variance. For scenarios (A), (B) and (C), 99\% of the variance resulted in 44, 77, and 119 components, respectively, extracted from 971,860 channel-wise examples.

A second dimensionality reduction is then performed ($\sigma()$), referred to as the filter function, with the goal of representing meaningful characteristics of the relationship between features~\citep{singh_originalmapper}. Within this study, \textit{t}-Stochastic Neighborhood Embedding (t-SNE)~\citep{original_tsne} is used to encapsulate important local structure between features. The two-dimensional (2D) t-SNE lens was constructed with a perplexity of 30, as this configuration resulted in the most stable visualization over many repetitions (tested on scenario (A)). Using t-SNE as part of the Mapper algorithm instead of on its own leverages its ability to represent local structure while avoiding the use of a low-dimensional manifold to encapsulate global structure. Instead, the global structure is predominantly incorporated into the topological network produced by Mapper during the fifth stage. 

The 2D lens was then segmented into a set of overlapped bins (the hypercubes centered on the elements of $\mathpzc{V}$), called the cover. A stable topological network was obtained when each dimension was divided into 5 regions, forming a grid of 25 cubes that were overlapped by 65\%. The number of regions correspond to the topological network's resolution, while the overlap has an influence on the amount of connection formed between nodes~\citep{singh_originalmapper}. 

Data points in each region are then clustered in isolation to provide insight into the local structure of the feature space (the elements of $\mathpzc{Z}_{\vec{v}}$ correspond to the data-point of a specific region).
For each region, Ward's hierarchical clustering ($\xi$) was applied to construct a dendogram that grouped similar features together according to a reduction in cluster variance~\citep{ward_originalclustering}.

Finally, the dendograms produced using neighboring regions are combined to form the topological network ($\mathpzc{G}$) using the features that lie in the overlapped area to construct the edges between the nodes.

The implementation of the Mapper algorithm was facilitated by a combination of the KeplerMapper~\citep{kmapper} and the DyNeuSR (Dynamical Neuroimaging Spatiotemporal Representations)~\citep{DyNeSR} Python modules. 
An extended coverage of processing pipelines for time-series TDA is given in \cite{PhinyomarkTDA2018}.


\vspace*{0.5cm}

\section{Results}
\label{results}

\subsection{Handcrafted features}
\label{sec:subsection_results_HF}
\begin{figure}[h!]
\begin{center}
\includegraphics[width=\textwidth]{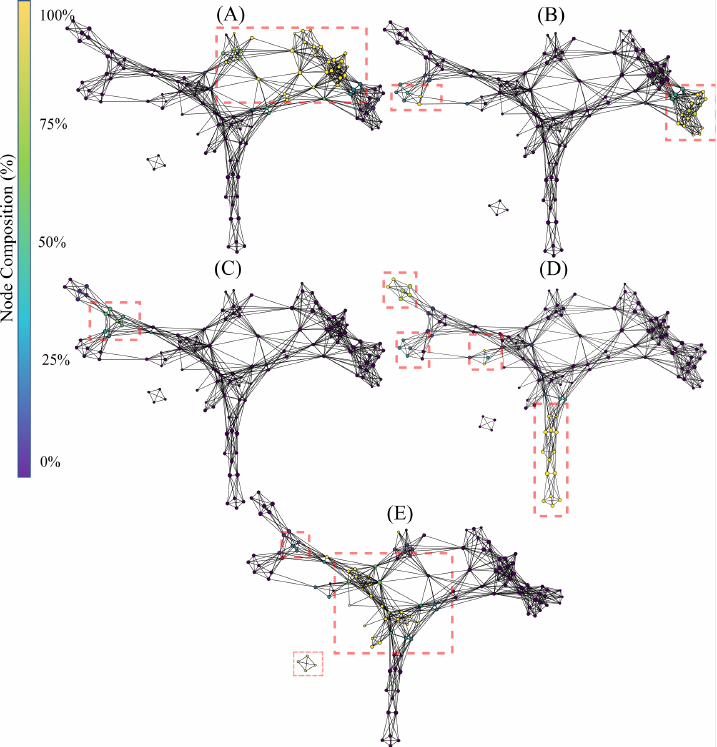}
\end{center}
\caption{Topological network generated exclusively for the handcrafted features, where nodes are colored to indicate percent composition of: (a) signal amplitude and power features (SAP), (b) nonlinear complexity (NLC), (c) frequency information features (FI), (d) time series modeling features (TSM), and (e) unique features (UNI). Dashed boxes highlight dense groupings of the specified functional group in each of the networks.}
\label{fig:handcraftedTN}
\end{figure}

Figure~\ref{fig:handcraftedTN} shows the topological network produced using only the handcrafted features.
The Kullback-Leibler divergence of the t-SNE embedding of the handcrafted features plateaued at 0.50, indicating that the perplexity and number of iterations used was appropriate for the dataset.
The topological network consisted of 125 nodes and 524 edges.

The color of the nodes within the network indicates the percentage of members that belong to the feature group of interest ((A):SAP, (B): NLC), (C): FI, (D): TSM, and (E): UNI).
The presence of an edge symbolizes common features present in the connected nodes,
which can be used at a global scale to verify that functional groups (similar information) cluster together.
Due to the topological nature of the graph, information similarity between nodes is measured using the number of edges that separate two nodes and not the length of the edges. Detailed interpretation of the TDA networks are given in the discussion.

\vspace*{0.5cm}

\subsection{Deep Features}

\begin{figure}[h]
    \includegraphics[width=\linewidth]{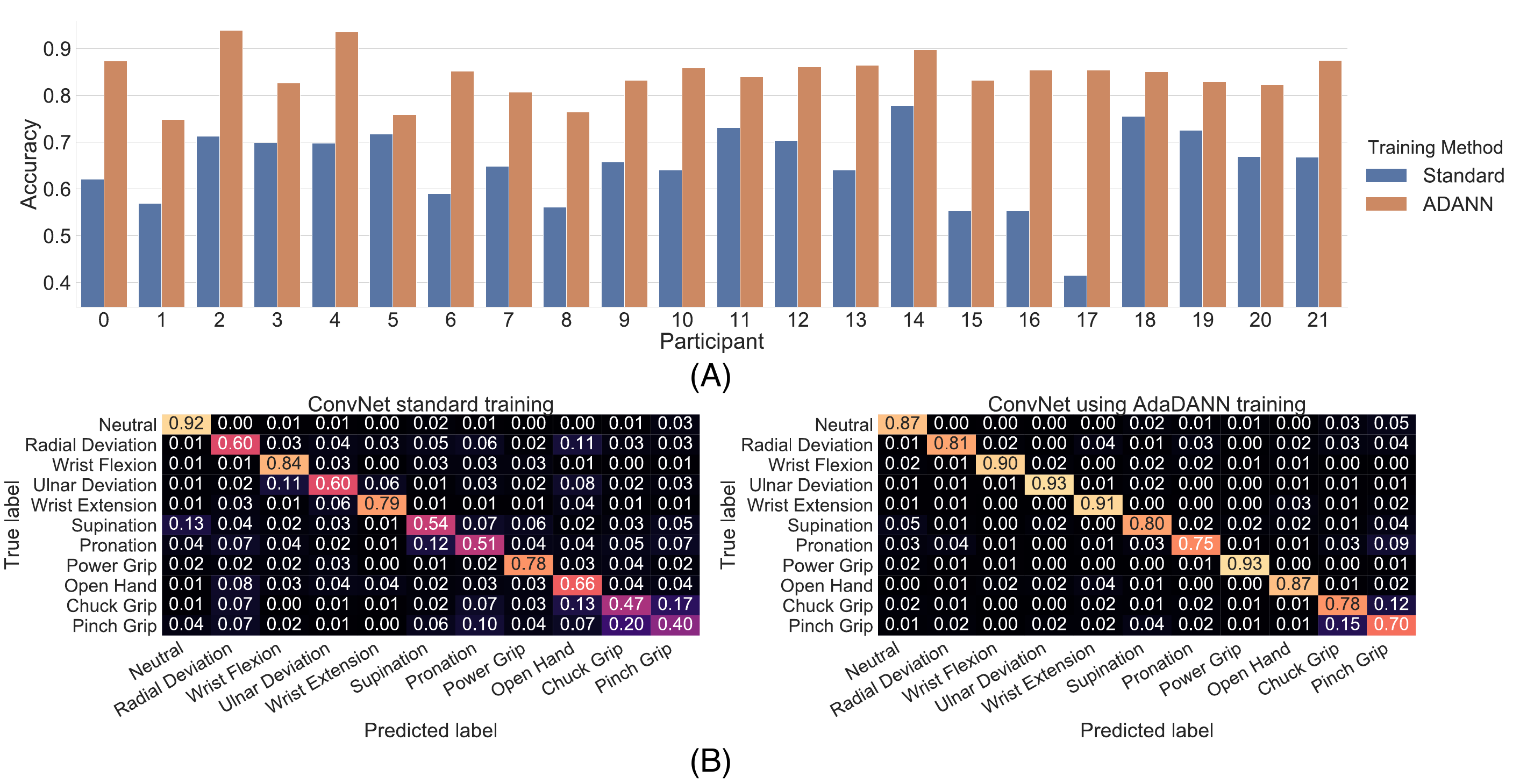}
    \caption{Classification results of deep learning architectures. A) Per-participant test set accuracy comparison when training the network with and without ADANN, B) Confusion matrices on the test set for cross-subject training with and without ADANN.}
        \label{fig:barplot_accuracy}
        \label{fig:confusioN_matrix_convnet}
\end{figure}

The average cross-subject accuracy on the test set when using the proposed ADANN framework was $84.43\%\pm 0.05\%$. Using a Wilcoxon signed-rank test~\citep{wilcoxon1992individual} with $n=22$, and considering each participant as a separate dataset, this was found to significantly outperform ($p<0.0001$) the average accuracy of $65.03\% \pm 0.08\%$ obtained when training the ConvNet conventionally.
Furthermore, based on Cohen's $d$, this difference in accuracy was considered to be huge~\citep{effect_size}. The accuracy obtained per participant for each training method is given in Figure~\ref{fig:barplot_accuracy}A, and the confusion matrices calculated on the gestures are shown in Figure~\ref{fig:confusioN_matrix_convnet}B.

\begin{figure}[t!]
    \centering
    \includegraphics[width=.85\textwidth]{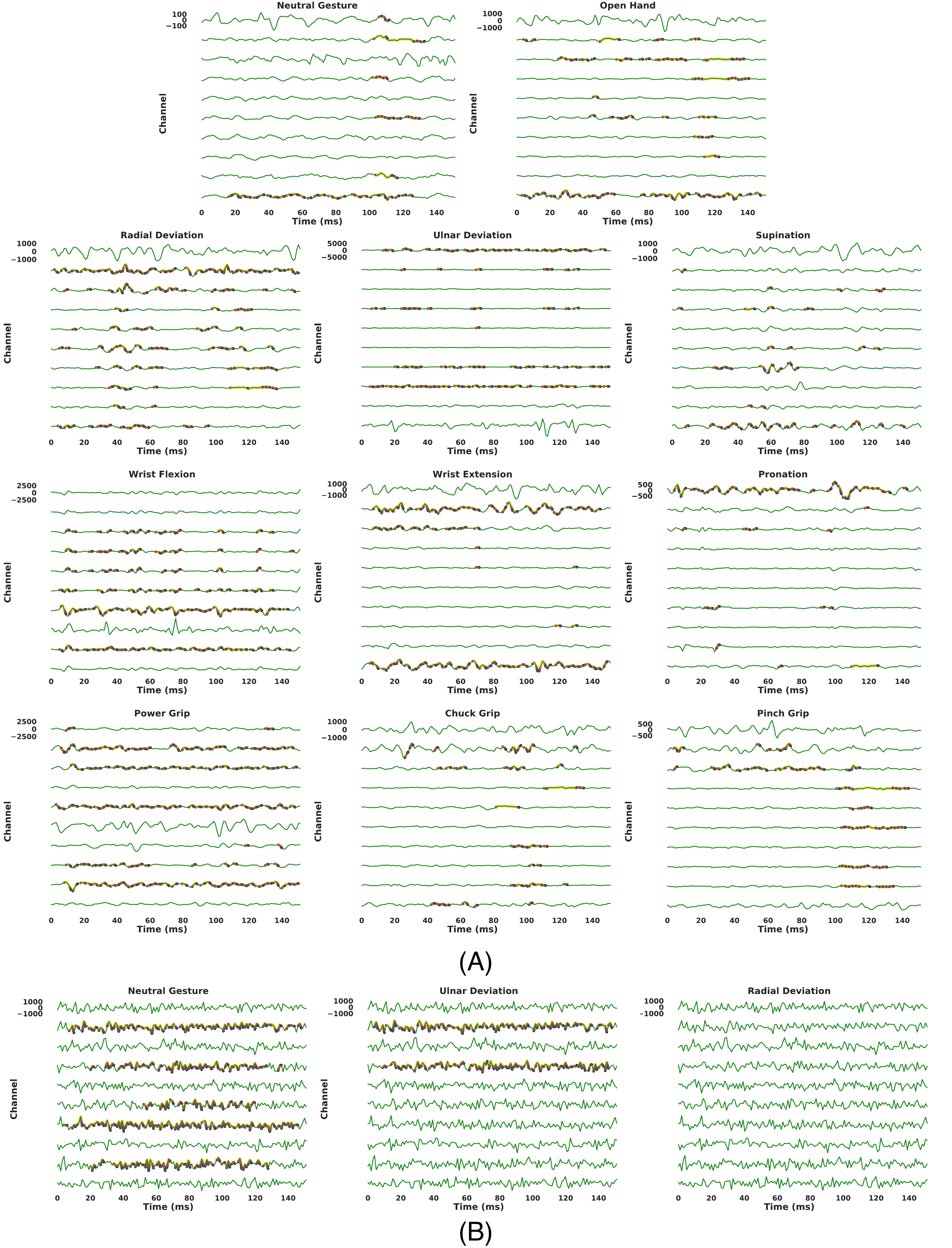}
    \caption{Output of Guided Grad-CAM when asked to highlight specific gestures in an example. For all graphs, the y-axis of each channel are scaled to the same range of value (indicated on the first channel of each graph). Warmer colors indicate a higher 'importance' of a feature in the input space for the requested gesture. The coloring use a logarithmic scale. For visualization purposes, only features that are within three order of magnitudes to the most contributing feature are colored. (A) The examples shown are real examples and correspond to the same gestures that Guided Grad-CAM is asked to highlight. (B) A single example, generated using Gaussian noise of mean 0 and standard deviation 450, is shown three times. While the visualization algorithm does highlight features in the input space (when the requested gesture is not truly present in the input), the magnitude of these contributions is substantially smaller (half or less) than when the requested gesture is present in the input.}
    \label{fig:combined_guided_grad_cam_results}
\end{figure}

Figure~\ref{fig:combined_guided_grad_cam_results}A provides visualizations of the ConvNet trained with ADANN using Guided Grad-CAM for several examples from the \textit{3DC Dataset},
These visualizations highlight what the network considers "important" (i.e., which part of the signals had the most impact in predicting a given class) for the prediction of a particular gesture.

\clearpage


Instead of using Guided Grad-CAM to visualize how the network arrived at a decision for a known gesture, Figure~\ref{fig:combined_guided_grad_cam_results}B presents the results of the visualization algorithm when the network is told to find a gesture that is not present in the input. This is akin to using a picture of a cat as an input to the network and displaying the parts of the image that most resemble a giraffe. In Figure~\ref{fig:combined_guided_grad_cam_results}B, the input was randomly generated from a Gaussian distribution of mean 0 and standard deviation of 450  (chosen to have the same scale as the EMG signals of the \textit{3DC Dataset}). For six of the eleven gestures (Radial Deviation, Wrist Extension, Supination, Open Hand, Chuck Grip and Pinch Grip) the network correctly identifies no relevant areas pertaining to these classes. While the network does highlight features in the input space associated with the other gestures, the magnitude of these contributions was substantially smaller (half or less) than when the requested gesture was actually present in the input signal.

\begin{figure}[h!]
\begin{center}
\includegraphics[width=0.90\textwidth]{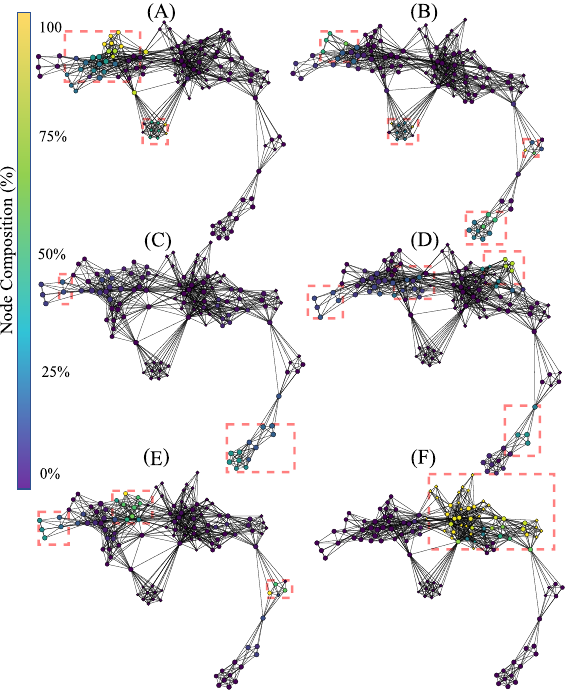}
\end{center}
\caption{Topological network generated for exclusively the learned features, where nodes are colored to indicate percent composition of: (A) Block 1's features, (B) Block 2's features, (C) Block 3's features, (D) Block 4's features, (E) Block 5's features, and (F) Block 6's features. Dashed boxes highlight dense groupings of the specified block features in each of the networks.}
\label{fig:learnedTN}
\end{figure}

The topological network produced using only the learned features is given in Figure~\ref{fig:learnedTN}. The color of the nodes within the network indicates the percentage of members that belong to the feature group of interests ((A): B1, (B): B2, (C): B3, (D): B4, (E): B5, and (F): B6).
Interpretation of the TDA network follows the rational stated in Section~\ref{sec:subsection_results_HF}.
The Kullback-Leibler divergence of the t-SNE embedding of the handcrafted features plateaued at 0.37, again indicating that the perplexity and number of iterations used was appropriate for the dataset.
The topological network consisted of 115 nodes and 672 edges.

\subsection{Hybrid Features}

\begin{figure}[h!]
\begin{center}
\includegraphics[width=\textwidth]{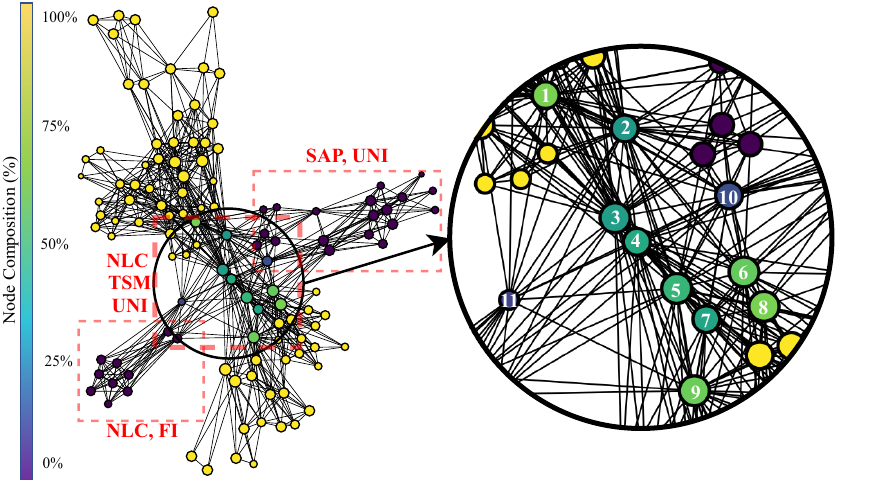}
\end{center}
\caption{Topological network generated for all features, where nodes were colored to indicate percent composition of learned features. The dashed boxes highlight dense grouping of handcrafted features with their associated type.} 
\label{fig:allTN}
\end{figure}

The topological network produced using both handcrafted and learned features is shown in Figure~\ref{fig:allTN}. 
The Kullback-Leibler divergence of the t-SNE embedding of all features plateaued at 0.53, again indicating that the perplexity and number of iterations used was appropriate for the dataset.
The topological network consisted of 115 nodes and 770 edges.
From this network, only a subset of nodes were occupied by both handcrafted and learned features.
Those nodes were indicated in Figure~\ref{fig:allTN}.

The color of the nodes within the network indicates the percentage of members that belong to the feature group of interests (learned features).
Information similarity was shown through a zoomed-in region of the network, where learned and handcrafted features clustered together.
The feature members of the numbered nodes were listed in Table \ref{table:allTN_features}.
Interpretation of the TDA network follows the rational stated in Section~\ref{sec:subsection_results_HF}.

\begin{table}[!htb]
\centering
\setlength\extrarowheight{10pt}{
\resizebox{\textwidth}{!}{%
\begin{tabular}{ccl}
\textbf{\#} &\textbf{Summary} & \textbf{Members}\\ \hline%

 \multirow{2}{*}{1} & \multirow{2}{*}{TSM+LeF5} & AR2 AR4 DAR2 DAR4 CC1 CC4 DCC1 \\
 & & DCC3 SNR 8xLeF1 1xLeF2 4xLeF4 10xLeF5  13xLe5\\ \hline
 
 \multirow{2}{*}{2} & \multirow{2}{*}{TSM+UNI+LeF6} & APEN AR2 AR4 DAR2 DAR4 CC1 CC4 DCC1 DCC3 DCC4 CE DFA DPR HIST123 \\
 & & SKEW MAVS OHM PSDFD PSR SMR SNR VCF VFD 1xLeF1 3xLeF2 3xLeF5 21xLeF6\\ \hline
 
 \multirow{2}{*}{3} & \multirow{2}{*}{TSM+UNI+LeF6} & APEN AR2 AR4 DAR2 DAR4 CC1 CC4 DCC1 DCC3 DCC4 CE DFA DPR HIST12 \\
 & & SKEW MAVS OHM PSDFD PSR SMR SNR VCF VFD 1xLeF1 1xLeF2 1xLeF5 27xLeF6\\ \hline
 
 \multirow{2}{*}{4} & \multirow{2}{*}{UNI+LeF6} & APEN DCC4 CE DFA DPR HIST123 \\
 & & SKEW MAVS OHM PSDFD PSR SMR VCF VFD 2xLeF2 2xLeF5 21xLeF6\\ \hline
 
 \multirow{2}{*}{2} & \multirow{2}{*}{TSM+UNI+LeF6} & APEN CC1 CC4 DCC4 CE DFA DPR HIST123  \\
 & & SKEW MAVS OHM PSDFD PSR SMR SNR VCF VFD 37xLeF6 \\ \hline
 
 \multirow{2}{*}{6} & \multirow{2}{*}{TSM+UNI+LeF6} & CC1 CC4 DCC4 CE DPR HIST123 SKEW MAVS PSDFD SMR  \\
 & & SNR VCF VFD 5xLeF2 5xLeF4 1xLeF5 37xLeF6\\ \hline
 
 \multirow{2}{*}{7} & \multirow{2}{*}{UNI+LeF6} & DCC4 CE DPR HIST123 SKEW MAVS  \\
 & & PSDFD SMR VCF VFD 2xLeF2 15xLeF6\\ \hline
 
 \multirow{2}{*}{8} & \multirow{2}{*}{UNI+LeF6} & DCC4 CE DPR HIST123 SKEW MAVS PSDFD SMR \\
 & & VCF VFD 5xLeF2 5xLeF4 1xLeF5 37xLeF6\\ \hline
 
 \multirow{2}{*}{9} & \multirow{2}{*}{UNI+LeF6} & APEN DCC4 CE DFA DPR HIST2 SKEW MAVS  \\
 & & OHM PSDFD PSR SMR VCF VFD 15xLeF2 36xLeF6\\ \hline
 
 \multirow{2}{*}{10} & \multirow{2}{*}{All Handcrafted+LeF6} & APEN CC14 DCC4 CE DFA DPR HIST123 KURT SKEW M2 MAVS MAX MHW23  \\ 
 & & MTW123 MNP TTP OHM PSDFD PSR SM SMR SNR SSI TM DTM VAR DVARV VCF VFD 11xLeF6\\\hline
 
 \multirow{2}{*}{11} & \multirow{2}{*}{NLC+LeF6} & APEN SAMPEN BC  \\
 & & KATZ 1xLeF6 \\ \hline
 
\end{tabular}}}
\caption{Members of nodes labeled in Figure~\ref{fig:handcraftedTN}. LeF\textbf{X} refers to a \textit{Learned Feature} from block \textbf{X}.}
\label{table:allTN_features}
\end{table}

\begin{figure}[!ht]
\begin{center}
\includegraphics[width=\textwidth]{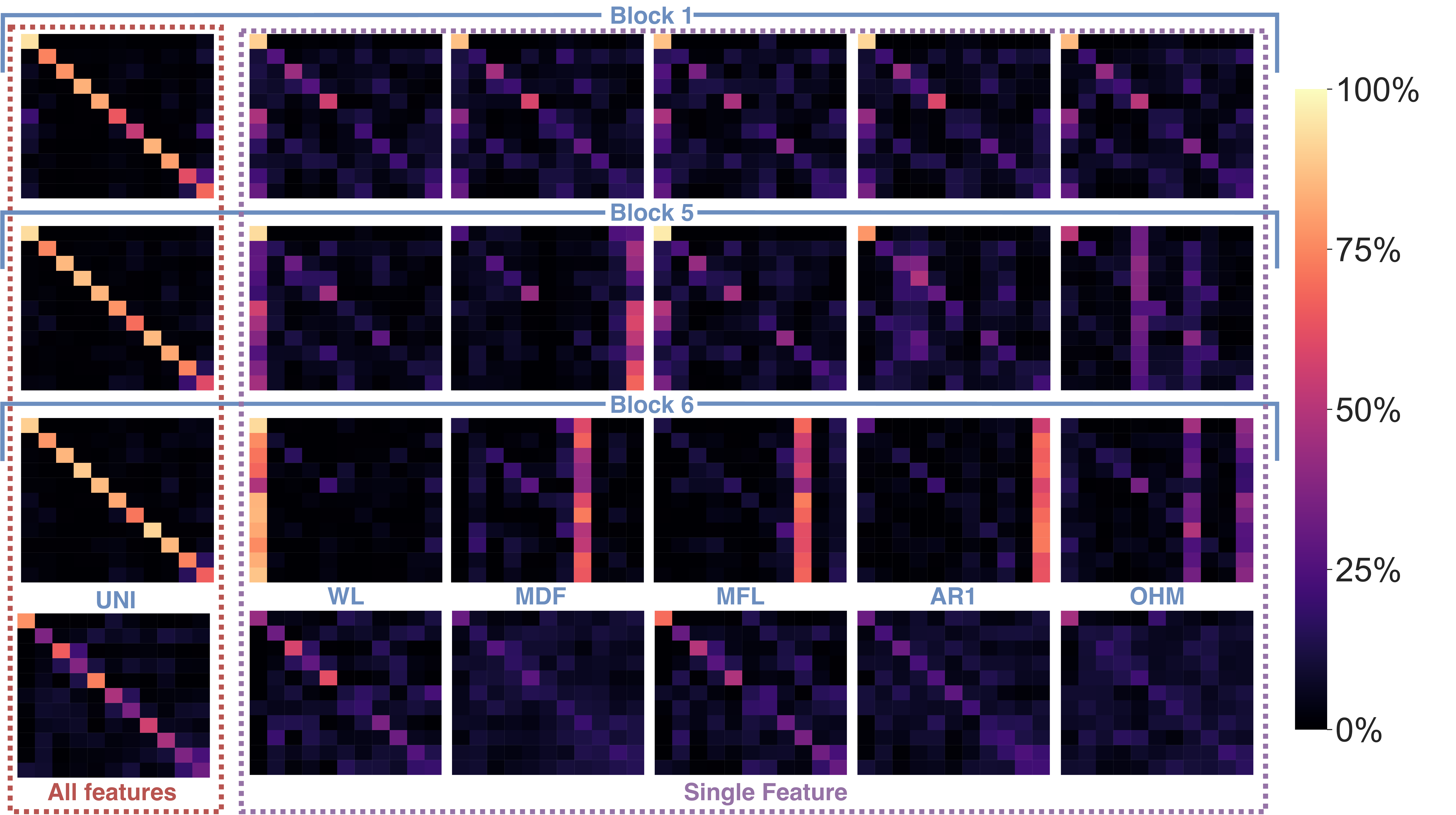}
\end{center}
\caption{Confusion matrices using the handcrafted features and the learned features from the first, penultimate and last block as input and a LDA as the classifier. The first column, denoted as \textit{All features}, shows the confusion matrices when using all 64 learned features of Block 1, 5 and 6 respectively (from top to bottom) and the set of UNI handcrafted features. The next five columns, denoted as \textit{Single Feature}, show the confusions matrices for handcrafted feature examplars and from the same network's blocks but when training the LDA on a single feature. The subset of learned features was selected as representative of the typical confusion matrices found at each block. The examplars of the handcrafted features were selected  from each handcrafted features' category (in order: SAP, FI, NLC, TSM and UNI). } 
\label{fig:lda_confusion_matrices}
\end{figure}

Table~\ref{Table_accuracies_per_block_lda} shows the average accuracy (grouped by block for the learned features and by group for the handcrafted features) obtained when training an LDA on each feature and when using all features within a category (i.e. within a block or within a group of handcrafted feature). Note that for the learned features, PCA is applied to the feature map and the first component is employed to represent a given learned feature. Figure~\ref{fig:lda_confusion_matrices} shows examples of confusion matrices computed from the LDA classifications of singular features (both handcrafted and learned). Figure~\ref{fig:lda_confusion_matrices}, also shows some confusion matrices obtained from the LDA's classification result when using all features within a category. 

\begin{table}[!ht]
\begin{center}
\begin{tabular}{@{}lcc|c@{}}
\toprule
 & \multicolumn{2}{c|}{Single Feature} & All Features \\
 & Average Accuracy & STD & Accuracy \\ \midrule
SAP & 26.80\% & \multicolumn{1}{c|}{7.0\%} &  41.61\% \\
FI & 19.95\% & \multicolumn{1}{c|}{2.87\%} &  34.80\% \\
NLC & 22.32\% & \multicolumn{1}{c|}{7.15\%} &  31.49\% \\
TSM & 22.24\% & \multicolumn{1}{c|}{3.33\%} &  37.18\% \\
UNI & 15.32\% & \multicolumn{1}{c|}{5.11\%} &  48.37\% \\
Block 1 & \textbf{28.49\%} & \multicolumn{1}{c|}{\textbf{3.84\%}} & 74.59\% \\
Block 2 & 28.28\% & \multicolumn{1}{c|}{4.66\%} & 78.26\% \\
Block 3 & 28.90\% & \multicolumn{1}{c|}{5.06\%} & 79.19\% \\
Block 4 & 29.21\% & \multicolumn{1}{c|}{5.15\%} & 78.77\% \\
Block 5 & 28.18\% & \multicolumn{1}{c|}{5.48\%} & 79.23\% \\
Block 6 & 26.62\% & \multicolumn{1}{c|}{6.19\%} & \textbf{81.38\%} 
\end{tabular}
\caption{Accuracy obtained on the test set using the handcrafted features and the learned features from their respective block. The \textit{Single Feature} accuracies are given as the average accuracy over all the features of their respective block/category.}
\label{Table_accuracies_per_block_lda}
\end{center}
\end{table}

\begin{figure}[!ht]
\begin{center}
\includegraphics[width=.8\textwidth]{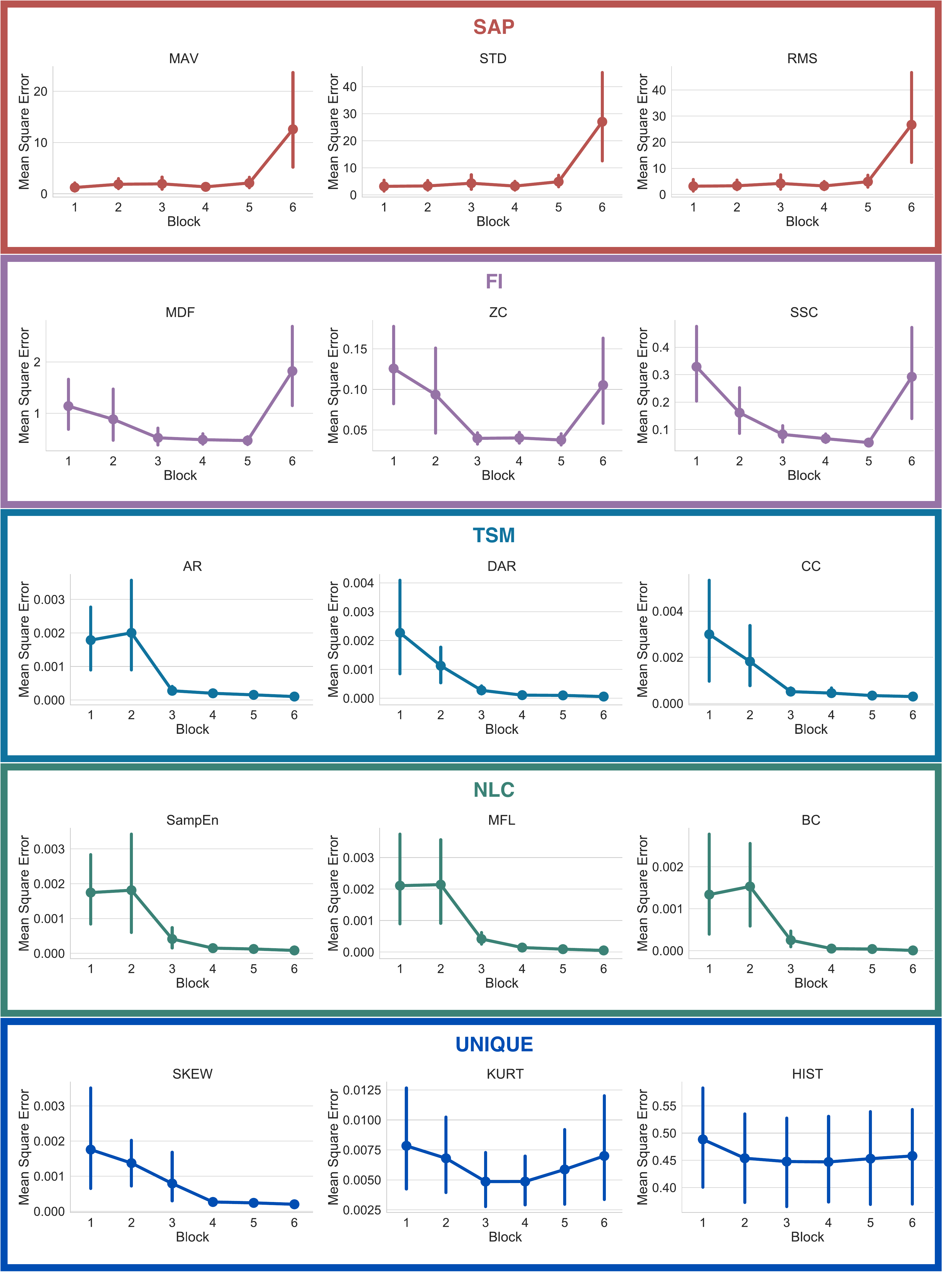}
\end{center}
\caption{Mean squared error of the regressions from learned features to handcrafted features, with respect to the number of blocks employed for the regression. The features are grouped with their respective functional groups.} 
\label{fig:featuresRegression}
\end{figure}

Figure~\ref{fig:featuresRegression} shows the average mean square error computed when regressing from the ConvNet's learned features (see Section~\ref{RegressionModelSection}) to fifteen handcrafted features (three per Functional Group). Note that the mean squared error is obtained by computing the regression using only the output of the block of interest.

\section{Discussion}
\label{discussion}

\subsection{Handcrafted Features}
The result of the Mapper algorithm applied to handcrafted features (see Figure~\ref{fig:handcraftedTN}) showed that the handcrafted features agglomerated mostly with their respective groups, and that the topological graph is Y-shaped. This shows that the hyperparameters selected in this work are consistent with those found in previous EMG literature~\citep{PhinyomarkTDA2018,evan_ner2019}.

\subsection{ADANN and Deep Learning Visualization}
Figure~\ref{fig:confusioN_matrix_convnet}B shows that training the network with ADANN outperforms the standard training method in cross-subject classification.
One advantage of ADANN in the context of this work is that the weights of the network have strong incentives to be subject-agnostic.
As such, the learned features extracted from the network can be thought of as general features (and to a certain extent subject-independent) for the task of sEMG-based hand gesture recognition.

Applying Guided Grad-CAM, as in \ref{fig:combined_guided_grad_cam_results}, 
shows that the network mostly focuses on different channels for the detection of antagonist gestures.
This suggests that the ConvNet was able to extract spatial features despite having access only to one dimensional convolutional kernels. Furthermore, it is notable that for all the examples given in Figure~\ref{fig:combined_guided_grad_cam_results}A, the most active channel was not the primary channel used for the gesture prediction. In fact, for the vast majority of gestures, the channel with the highest amplitude did not contribute in a meaningful way to the network's prediction. This observation held true while looking at several other examples from the \textit{3DC Dataset}. This might indicate that the common practice of placing the recording channel directly on the most prominent muscle for a given gesture within the context of gesture recognition may not be optimal. One could thus use the type of information provided by algorithms such as Guided Grad-CAM as another way of performing channel selection (instead of simply using classification accuracy). 
The absence of importance on amplitude characteristics is in contrast to conventional practices of handcrafted feature engineering - where the feature set typically relies heavily on amplitude characteristics.
This perhaps explains the growing interest in handcrafted feature extraction techniques that do not capture amplitude information, such as TDPSD, that have been demonstrated to outperform conventional amplitude-reliant features in terms of accuracy and robustness to confounding factors~\citep{Khushaba2016}.

When applying Guided Grad-CAM on a noise input (one where the target gesture is not present, as seen in Figure~\ref{fig:combined_guided_grad_cam_results}B), the reported activation level is substantially lower, and in some cases nonexistent. When the standard deviation of the Gaussian noise was increased by 33\%, the network did not find any features resembling any gesture. This is most likely due to the fact that increasing the spread of the noise leads to a potentially greater gap in value between two adjacent data-points (reduced smoothness) fostering the condition for a more unrealistic signal. One could thus imagine training a generative adversarial network with the discriminative function based on the activation level calculated by Guided Grad-CAM, and modulating the difficulty by augmenting the signal's amplitude. This could facilitate training a network to not only be able to generate realistic, synthetic EMG signal, but also have the signal resemble actual gestures. 


In contrast to the topological networks based on handcrafted features,  those based on the learned features appear as a long flair with a loop.
From Figure~\ref{fig:learnedTN}a, the learned features from block 1 are concentrated in the left segment of the flare, and the lower segment of the loop.
From Figure~\ref{fig:learnedTN}b, the learned features from block 2 were located slightly more central to the network than the block 1 features.
Additionally, a small subset of block 2 features appeared at the right segment of the flare, indicating a second distinct source of information was being harnessed.
From Figure~\ref{fig:learnedTN}c, d, and e, the features of block 3, 4, and 5 relocate their concentration of features to converge in the center of the network.
Finally from Figure~\ref{fig:learnedTN}f, the concentration of all block 6 features lies in the center of the network.
Thus, it can be seen that learned features from the same block tend to cluster together and remain close in the map to adjacent blocks in the network. The only exception to this is from the first block to the second, where substantially different features were generated by the latter. This suggests that the first layer may serve almost as a preprocessing layer which conditions the signal for the other layers.

\subsection{Hybrid Features Visualization}

The topological network generated from using both the handcrafted and learned features (see Figure~\ref{fig:allTN}) followed two orthogonal axes with the handcrafted features on one and the learned features on the other. The middle of the graph (where the two axis intercept) is where any nodes containing both handcrafted and learned features are found. The vast majority of these nodes are populated by features from block 6 and the NLC, TSM and UNI functional groupings. No nodes in the graph contained both handcrafted features and features from block 3, suggesting that block 3 extracted features not captured by current feature designs. Conversely, no learned features shared a node with features from the FI family, suggesting that these features may not have been extracted by the network.

While this topological network informs the type of information encoded within each individual feature, it is important to note that information can still be present but encoded in a more complex way within the weights of the deep network. This information flow can be visualized from the regression graphs of Figure~\ref{fig:featuresRegression}. Features from the SAP family are more easily predicted within the early blocks whereas features from the TSM and NLC family require the latter blocks of the network to achieve the best predictions. Interestingly, while features from the FI family did not share any nodes learned features, one can see that the deep network is able to better extract this type of information within the intermediary blocks. This indicates (from Figure~~\ref{fig:allTN}, \ref{fig:featuresRegression}) that, while frequency information is not explicitly used by the ConvNet, this type of information is nonetheless indirectly used to compute the features from the latter blocks.  An example of a feature for which the ConvNet was unable to leverage its topology is the HIST (see Figure~\ref{fig:featuresRegression}). 

\vspace{1cm}
\subsection{Understanding deep features predictions}

The topological network of Figure~\ref{fig:allTN} showed that the type of information encoded within the lower blocks of the ConvNet tended to be highly dissimilar to what the handcrafted features encoded. Interestingly, however, Figure~\ref{fig:lda_confusion_matrices} shows that the role fulfilled by these features is similar. That is, both the handcrafted and learned features (from the lower blocks) try to encode general properties that can distinguish between all classes. The confusion matrices obtained from training an LDA on a single feature highlight this behavior (see Figure~\ref{fig:lda_confusion_matrices} for some examples) as both the handcrafted features and the learned features (before the last block) are able to distinguish between gestures relatively equally. In contrast, the features extracted from the last block (and to a lesser extent from the penultimate block) have been optimized to be a gesture detector instead of a feature detector. A clear visual of this behavior is illustrated in Figure~\ref{fig:lda_confusion_matrices}, where the main line highlighted in the confusion matrices from block 6 was a single column (corresponding to the prediction of a single gesture), instead of the typical diagonal. In other words, during training, the neurons of the final block are encoded to have maximum activation when a particular class was provided in the input window and minimum activation when other classes were provided; effectively creating a one-versus-all (OVA) classifier. This behavior is consistent with the feature visualization literature found in image classification and natural language processing, where semantic dictionaries or saliency maps have depicted neuron representations becoming more abstract at later layers~\citep{saliancy_maps, deep_learning_presentation}. This also explains why the features from the last block obtained the worst average accuracy when taken individually while achieving the highest accuracy as a group (see Table~\ref{Table_accuracies_per_block_lda}). That is, as each feature map of the last layer tries to detect a particular gesture, its activation for the other gestures should be minimal, making the distinction between the other gestures significantly harder. The final decision layer of the network can then be thought of as a weighted average of these OVA classifiers to maximize the performance of the learned feature maps. Note that in Table~\ref{Table_accuracies_per_block_lda}, the lower accuracies obtained from the handcrafted features as a group were expected as each feature within the same family provides similar type of information, even more so than the learned features of the network (as seen in Figure~\ref{fig:handcraftedTN},~\ref{fig:learnedTN},~\ref{fig:allTN}). Overall, the best performing handcrafted feature set as a group was the features from the UNI family despite the fact that they were the worst on average when alone. This is most likely due to the fact that by definitions, features within this family are more heterogeneous. 

\section{Conclusion}
\label{conclusion}
This paper presents the first in-depth analysis of features learned using deep learning for EMG-based hand gesture recognition. The type of information encoded within learned features and their relationship to handcrafted features were characterized employing a mixture of topological data analysis (Mapper), network interpretability visualization (Guided Grad-CAM), machine learning (feature classification prediction) and by visualizing the information flow using feature regression. 
As a secondary, but significant contribution, this work presented ADANN, a novel multi-domain training algorithm particularly suited for EMG-based gesture recognition shown to significantly outperform traditional training on cross-subject classification accuracy.

This manuscript paves the way for hybrid classifiers that contain both learned and handcrafted features.
An ideal application for the findings of this work would rely on a mix of handcrafted features and learned features taken from all four extremities of the hybrid topological network, and at the center to provide complementary, and general features to the classifier. A network could then be trained to augment its sensitivity to similar classes.
For example, to alleviate ambiguity between pinch grip and chuck grip, a learned feature that encodes the one-versus-all information of pinch grip could be included into the original feature set or into an otherwise handcrafted only feature set.
Alternatively, handcrafted feature extraction stages may be installed within the deep learning architecture by means of neuroevolution of augmenting topologies~\citep{chen_neat}, a genetic algorithm that optimizes the weights and connections of deep learning architectures.

The main limitation of this study was the use of a single architecture to generate the learned features. Though this architecture was chosen to be representative of current practices in myoelectric control and be extensible to other applications, the current work did study the impact of varying the number of blocks and the composition of these block on the different experiments. Additionally, although the set of handcrafted features was selected to be comprehensive over the sources of information available from the EMG signal, explicit time-frequency features such as those based on spectrograms and wavelet were not included in the current work, as they were ill-adapted to the framework employed in this study. Furthermore, an analysis including a larger amount of gestures should also be conducted. Importantly, these results are presented for a single 1D electrode array, and may not be representative of larger 2D arrays such as those used in high density EMG applications.
Similarly, explicit spatio-temporal features, such as coherence between electrodes, were not explored, and the convolutional kernels were restricted to 1D (although as seen in Figure~\ref{fig:combined_guided_grad_cam_results}A the network was still able to learn spatial information to a certain extent). Omitting these type of complex features was a design choice as this work represents a first step in understanding and characterizing learned features within the context of EMG signal. 
As such, using this manuscript as a basis, future works should study the impact of diverse architectures on the type of learned features and will incorporate spatio-temporal features (both handcrafted and from 2D convolutional kernels). Additionally, formal feature set generation and hybrid classifiers should be investigated using the tools presented in this work.

\section*{Conflict of Interest Statement}
The authors declare no conflict of interest. The funders had no role in the design of the study; in the collection, analyses, or interpretation of the data; in the writing of the manuscript; or in the decision to publish the results.

\section*{Author Contributions}
Conceptualization: UCA, EC, AP, FL, BG and ES; methodology, UCA, EC, AP, FL and ES; software, UCA and EC; validation, UCA and EC; formal analysis, UCA, EC, AP, FL and ES; investigation, UCA, EC, AP, FL and ES; resources, FL, BG, ES; data curation, UCA and EC; writing—original draft preparation, UCA and EC; writing—review and editing, UCA, EC, AP, FL, BG and ES; visualization, UCA and EC; supervision, FL, BG and ES; project administration, UCA, EC, AP, FL, BG and ES; funding acquisition, UCA, AP, FL, BG and ES;

\section*{Funding}
This research was funded by the Natural Sciences and Engineering Research Council of Canada (NSERC) [funding reference numbers 401220434, 376091307, 114090], the Institut de recherche Robert-Sauvé en santé et en sécurité du travail (IRSST) and the Canada Research Chair in Smart Biomedical Microsystems [funding reference number 950-232064]. Cette recherche a été financée par le Conseil de recherches en sciences naturelles et en génie du Canada (CRSNG) [numéros de référence 401220434, 376091307, 114090].

\section*{Acknowledgments}
The authors would like to thank Gabriel Dubé for his valuable input in relation to the Mapper algorithm. 


\section*{Data Availability Statement}
The datasets analyzed and the source code for this study can be found at the following link:\\ \href{https://github.com/UlysseCoteAllard/sEMG\_handCraftedVsLearnedFeatures}{https://github.com/UlysseCoteAllard/sEMG\_handCraftedVsLearnedFeatures}.

\bibliographystyle{formatting/frontiersinSCNS_ENG_HUMS} 
\bibliography{frontiers}





\end{document}